\documentclass[twocolumn]{aastex6}

\graphicspath{{./figures/}}
\usepackage{amsmath,amssymb}
\usepackage{comment}
\usepackage{url}
\def\bm#1{\mbox{\boldmath $#1$}}

\renewcommand{\b}{$\mathrm{b}$}

\newcommand{\p}{\mathrm{p}}
\newcommand{\jup}{\mathrm{Jup}}
\newcommand{\kbjd}{\mathrm{BJD_{TDB}}-2454833}
\newcommand{\gcc}{\mathrm{g\,cm^{-3}}}
\newcommand{\days}{\mathrm{days}}
\newcommand{\uni}{\mathcal{U}}
\newcommand{\luni}{\mathcal{U}_{\rm log}}
\newcommand{\gaus}{\mathcal{G}}
\newcommand{\red}{\textcolor{black}}

\shortauthors{Masuda}

\begin{document}


\title{
Eccentric Companions to Kepler-448\b\ and Kepler-693\b:\\
Clues to the Formation of Warm Jupiters
}

\email{kmasuda@astro.princeton.edu}

\author{Kento Masuda\altaffilmark{1,2,3}}
\altaffiltext{1}{Department of Astrophysical Sciences, Princeton University, Princeton, NJ 08544, USA} 
\altaffiltext{2}{Department of Physics, The University of Tokyo, Tokyo 113-0033, Japan}
\altaffiltext{3}{NASA Sagan Fellow}

\slugcomment{
Accepted for publication in AJ on June 18}

\begin{abstract}

\red{I report the discovery of non-transiting close companions to two transiting warm Jupiters (WJs), 
Kepler-448/KOI-12b 
(orbital period $P=17.9\,\days$, radius $R_{\rm p}=1.23^{+0.06}_{-0.05}\,R_{\rm Jup}$)
and Kepler-693/KOI-824b 
($P=15.4\,\days$, $R_{\rm p}=0.91\pm0.05\,R_{\rm Jup}$),
via dynamical modeling of their transit timing and duration variations (TTVs and TDVs). The companions have masses of $22^{+7}_{-5}\,M_\jup$ (Kepler-448c) and $150^{+60}_{-40}\,M_\jup$ (Kepler-693c), and both are on eccentric orbits ($e=0.65^{+0.13}_{-0.09}$ for Kepler-448c and $e=0.47^{+0.11}_{-0.06}$ for Kepler-693c) with periastron distances of $1.5\,\mathrm{au}$. Moderate eccentricities are detected for the inner orbits as well ($e=0.34^{+0.08}_{-0.07}$ for Kepler-448b and $e=0.2^{+0.2}_{-0.1}$ for Kepler-693b). 
In the Kepler-693 system, a large mutual inclination between the inner and outer orbits ($53^{+7}_{-9}\,\mathrm{deg}$ or $134^{+11}_{-10}\,\mathrm{deg}$) is also revealed by the TDVs. This is likely to induce a secular oscillation of the inner WJ's eccentricity that brings its periastron close enough to the host star for tidal star--planet interactions to be significant. In the Kepler-448 system, the mutual inclination is weakly constrained and such an eccentricity oscillation is possible for a fraction of the solutions. Thus these WJs may be undergoing tidal migration to become hot Jupiters (HJs), although the migration via this process from beyond the snow line is disfavored by the close-in and massive nature of the companions. 
This may indicate that WJs can be formed in situ and could even evolve into HJs via high-eccentricity migration inside the snow line.
}

\end{abstract}

\keywords{planets and satellites: formation 
--- planets and satellites: individual (KOI-12, Kepler-448, KIC 5812701)
--- planets and satellites: individual (KOI-824, Kepler-693, KIC 5164255)
--- techniques: photometric
}



\section{Introduction}\label{sec:intro}

Warm Jupiters (WJs), giant planets in moderately close-in orbits ($7\,\days<P<100\,\days$), pose a similar conundrum to that of hot Jupiters (HJs). 
A dozen WJs have been found to reside in circular orbits in multi-transiting systems \citep{2016ApJ...825...98H}, in which the orbital planes of the planets are likely well aligned. Such an architecture points to the formation via disk migration \citep{1980ApJ...241..425G, 1996Natur.380..606L} as originally proposed for HJs, or in situ formation inside the snow line \citep{2016ApJ...817L..17B, 2016ApJ...829..114B}. Alignments of the planetary orbits with their host stars' equators, as confirmed for some of them \citep[e.g.,][]{2012Natur.487..449S, 2012ApJ...759L..36H}, may also support the absence of past 
dynamical eccentricity/inclination excitation via planet--planet scattering \citep{1996Sci...274..954R} or secular chaos \citep{2011ApJ...735..109W}. 

On the other hand, roughly half of WJs with measured masses from Doppler surveys have significant eccentricities that seem too large to result from disk migration or subsequent planet--planet scattering \citep{2014ApJ...786..101P}, but yet too small for their orbits being tidally circularized \citep{2012ApJ...750..106S, 2015ApJ...798...66D}. 
A possible explanation is that those moderately eccentric WJs are experiencing ``slow Kozai--Lidov migration" \citep{2015ApJ...799...27P}: their eccentricities are still undergoing large oscillations driven by the secular perturbation from a close companion \citep{2014ApJ...781L...5D}, without being suppressed by other short-range forces \citep{2003ApJ...589..605W}, and their orbits shrink only at the high-eccentricity phase. This scenario may indeed reproduce the observed eccentricity distribution of WJs with outer companions \citep{2016ApJ...829..132P}.

Observationally, long-period, massive companions to WJs are nearly as common as those of HJs \citep{2014ApJ...785..126K} and their orbital properties might be consistent with what is expected from this scenario \citep{2016ApJ...821...89B}. Indeed, the apsidal misalignments of some of those eccentric WJs with outer companions provide statistical evidence for the oscillating eccentricity due to a large mutual orbital inclination \citep{2014Sci...346..212D}. 
However, there has been no direct measurement of such a large orbital misalignment between WJs and their outer companions as to induce a large eccentricity oscillation, partly because they are mostly the systems detected with the radial velocity (RV) technique that yields no information on the orbital direction. 
Notable exceptions include the Kepler-419 system \citep{2014ApJ...791...89D} and the doubly-transiting giant-planet system Kepler-108 \citep{2017AJ....153...45M}, where the mutual inclinations were constrained via dynamical modeling of transit timing and duration variations (TTVs and TDVs),
though their mutual inclinations are likely too small to drive secular eccentricity oscillations.

Transiting WJs without transiting companions provide a unique opportunity to search for close and mutually-inclined companions as direct evidence for the slow Kozai migration, because the full 3D architecture of the system can be dynamically constrained 
\red{with a similar technique as used in the above systems.}
In addition, WJs on eccentric and intermediate orbits, unlike HJs, may still be interacting with the companion strongly, and their eccentricity 
can also help the TTV inversion for non-transiting objects by producing specific non-sinusoidal features. The TTV search for the outer companions on such ``hierarchical" orbits is also complementary to the TTV studies so far, which have mainly focused on nearly sinusoidal signals typical for planets near mean-motion resonances \citep{2016arXiv161103516H, 2016ApJ...820...39J}. 

In this paper, I perform a search for non-transiting companions around transiting WJs using transit variations (Section \ref{sec:targets}) and report the discovery of outer companions to two transiting WJs, Kepler-448b \citep{2015A&A...579A..55B} and Kepler-693b \citep{2016ApJ...822...86M}. Based on TTVs and TDVs of the WJs, I find that the companions are (sub-)stellar mass objects on highly eccentric, au-scale orbits (Sections \ref{sec:method}--\ref{sec:koi824}). In particular, I confirm a large mutual orbital inclination between the inner WJ and the companion in the latter system, which can induce a large amplitude of eccentricity oscillation and the tidal shrinkage of the inner orbit (Section \ref{sec:secular}) --- exactly as predicted in the ``slow Kozai" scenario by a close companion. \red{The companions' properties, however, are not fully compatible with such migration starting from beyond the snow line. Thus I also assess the possibility of in situ formation (Section \ref{ssec:secular_insitu}). I discuss possible follow-up observations in Section \ref{sec:discussions} and Section \ref{sec:summary} concludes the paper.}

\newpage


\section{Systematic TTV Search for Singly-Transiting Warm Jupiters}\label{sec:targets}

To identify the signature of outer companions, I analyzed TTVs of 
$23$ confirmed, singly-transiting WJs (Section \ref{ssec:targets_ttv}) with the orbital period $7\,\days<P<100\,\days$ and radius $R_\p>8R_\oplus$ in the DR24 of the KOI catalog \citep{2016ApJS..224...12C}. Systems with multiple KOIs are all excluded, even though they consist of only one confirmed planet and false positives. 
I found clearly non-sinusoidal TTVs for Kepler-448/KOI-12b, Kepler-693/KOI-824b, and Kepler-419/KOI-1474b. The result is consistent with the TTV search by \citet{2016ApJS..225....9H}, who reported significant long-term TTVs for the same three KOIs in our sample.\footnote{Due to the update in the stellar radius in the DR25 catalog, the WJ sample defined above now includes Kepler-522/KOI-318b and Kepler-827/KOI-1355b, for which \citet{2016ApJS..225....9H} reported long-term TTVs. Interpretation of their TTVs is less clear than for the two systems discussed in the present paper, because of the lack of clear non-sinusoidal features (Kepler-522b) and a low signal-to-noise ratio (Kepler-827b).}

Among those planets, Kepler-419b's TTVs have already been analyzed by \citet{2014ApJ...791...89D} with RV data, and the companion planet Kepler-419c was found to be a super Jupiter on an eccentric and mutually-aligned orbit with the inner one. Therefore in this paper, I focus on Kepler-448b and Kepler-693b, which both show clear non-sinusoidal TTVs, and masses and orbits of the perturbers can be well constrained. 

\begin{figure*}
	\centering
	\fig{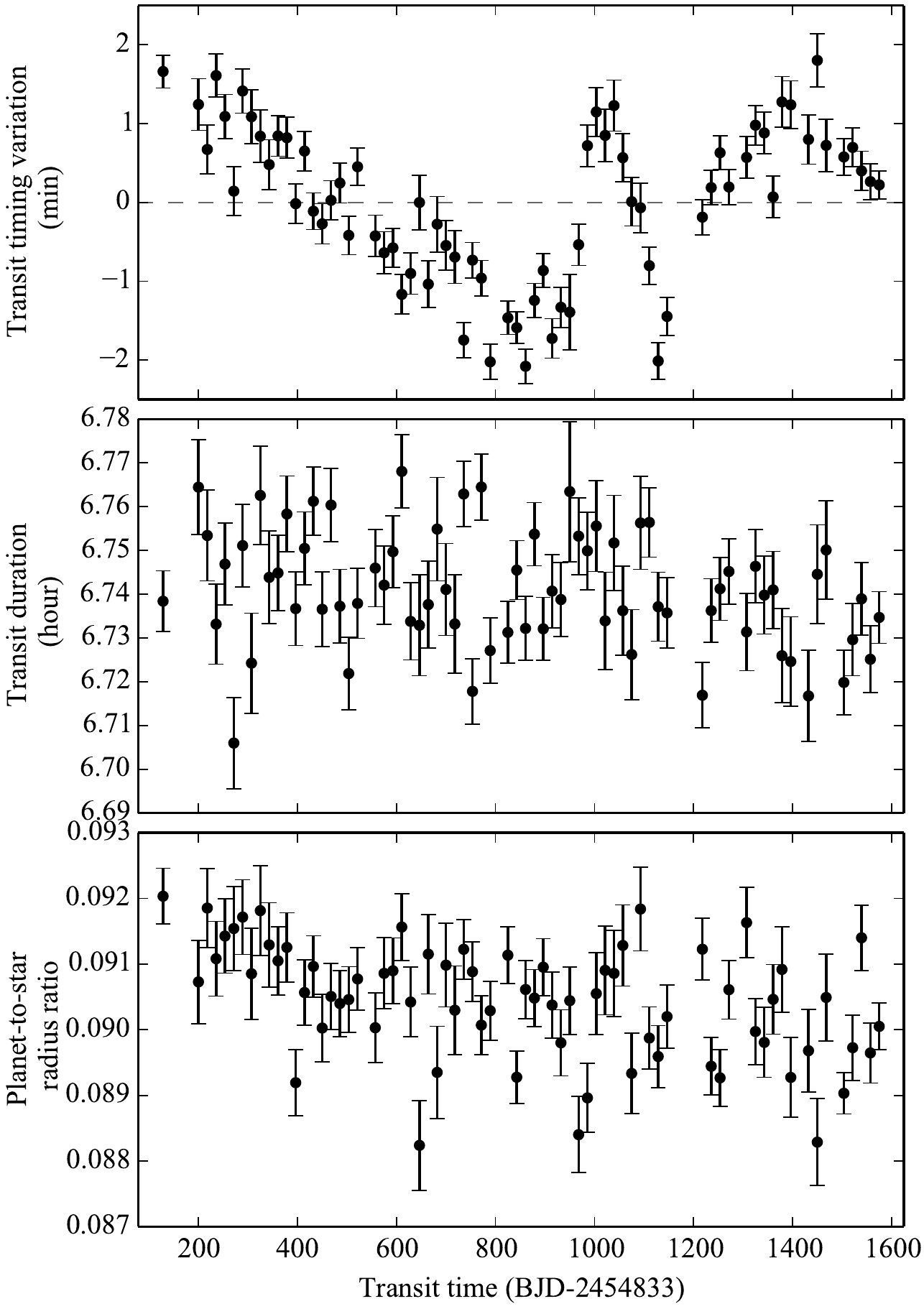}{0.48\textwidth}{(a) Kepler-448b}
	\fig{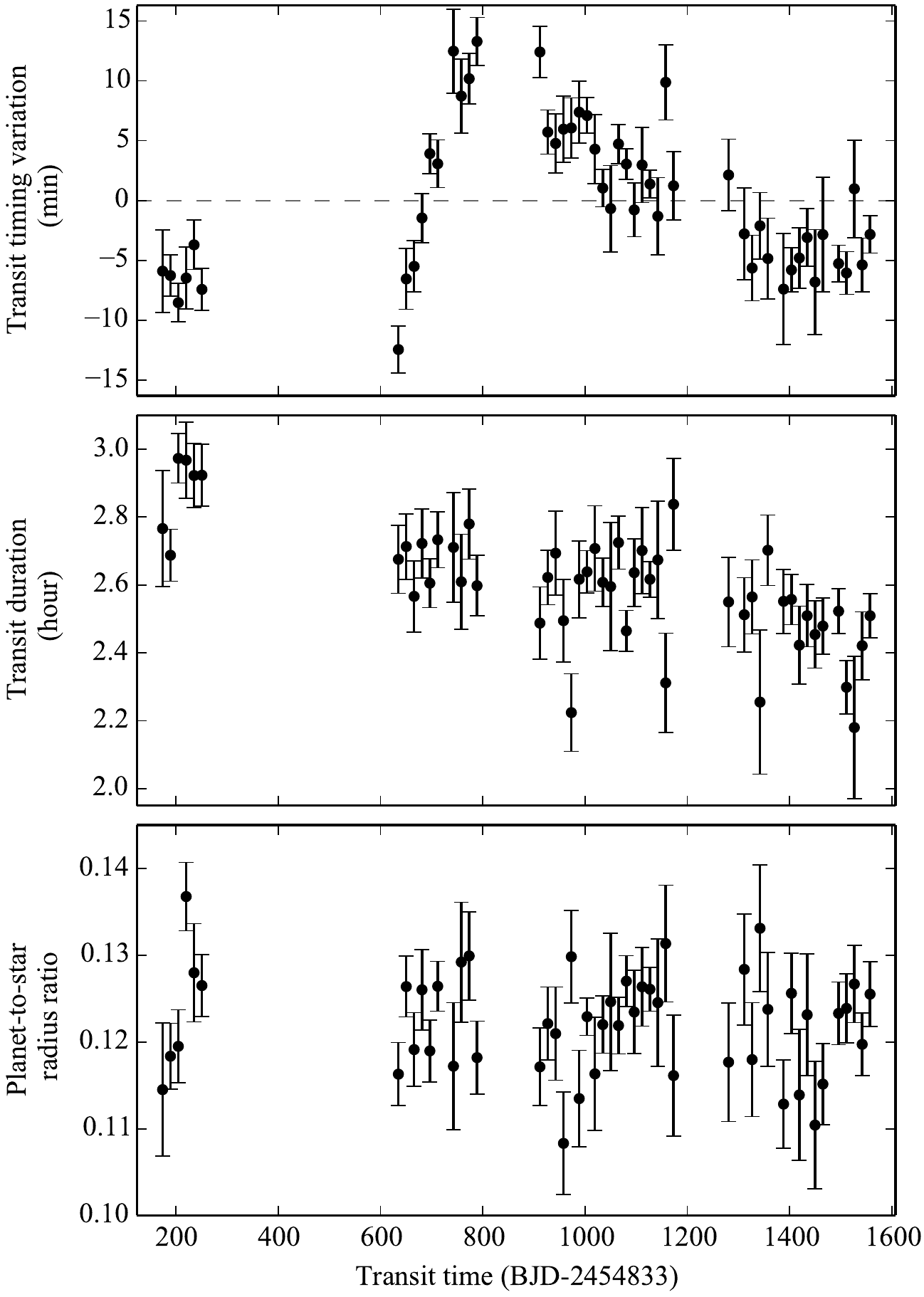}{0.48\textwidth}{(b) Kepler-693b}
	\caption{Central times, durations, and planet-to-star radius ratios
	obtained by fitting each transit of (a) Kepler-448b (Table \ref{tab:tc_dur_rp_koi12})
	and (b) Kepler-693b (Table \ref{tab:tc_dur_rp_koi824}).
	In the top panel, residuals from the linear fit to the observed transit times
	(i.e., TTVs) are shown for clarity. 
	}
	\label{fig:tc_dur_rp}
\end{figure*}

\begin{figure*}
	\centering
	\fig{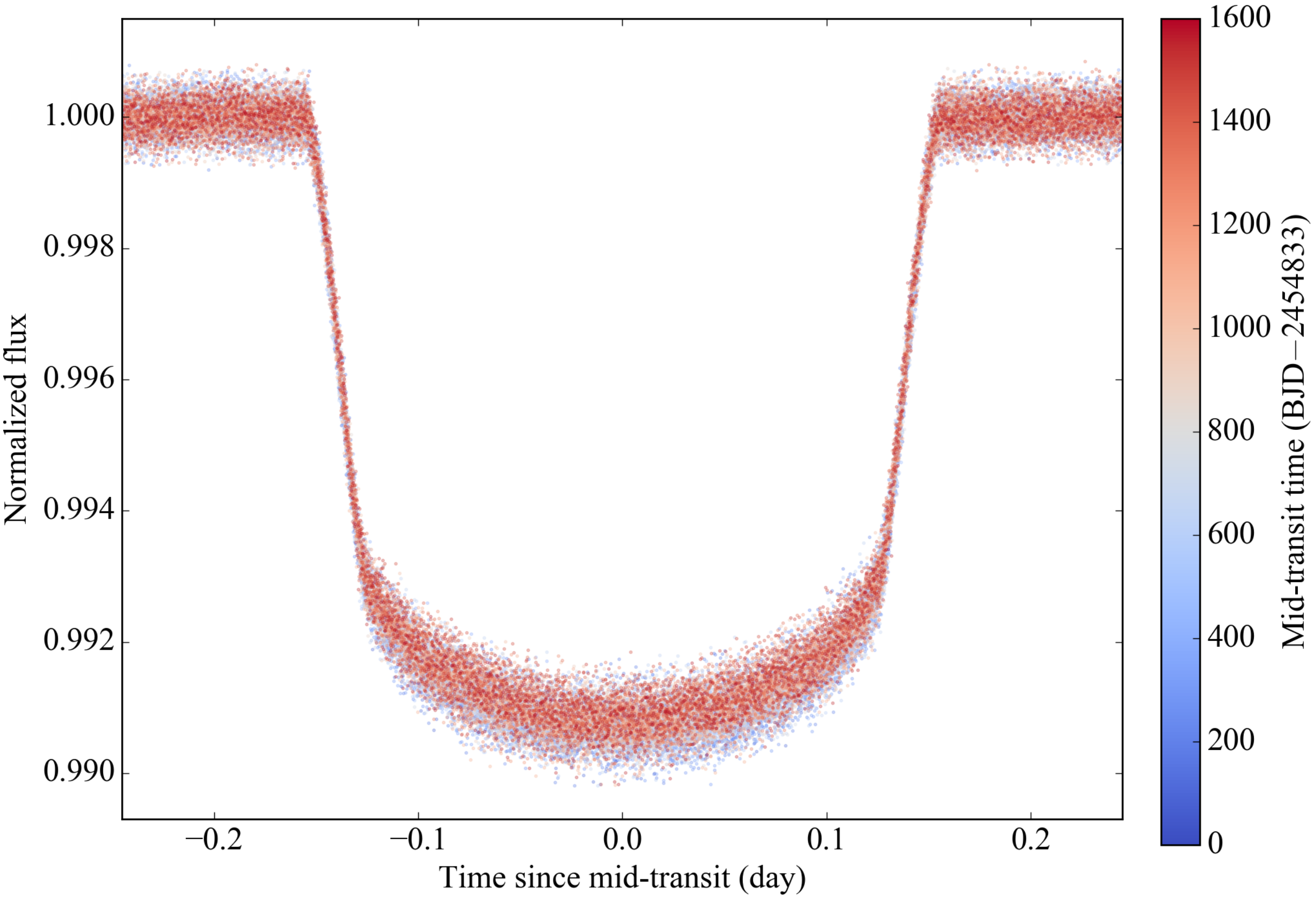}{0.49\textwidth}{(a) Kepler-448b}
	\fig{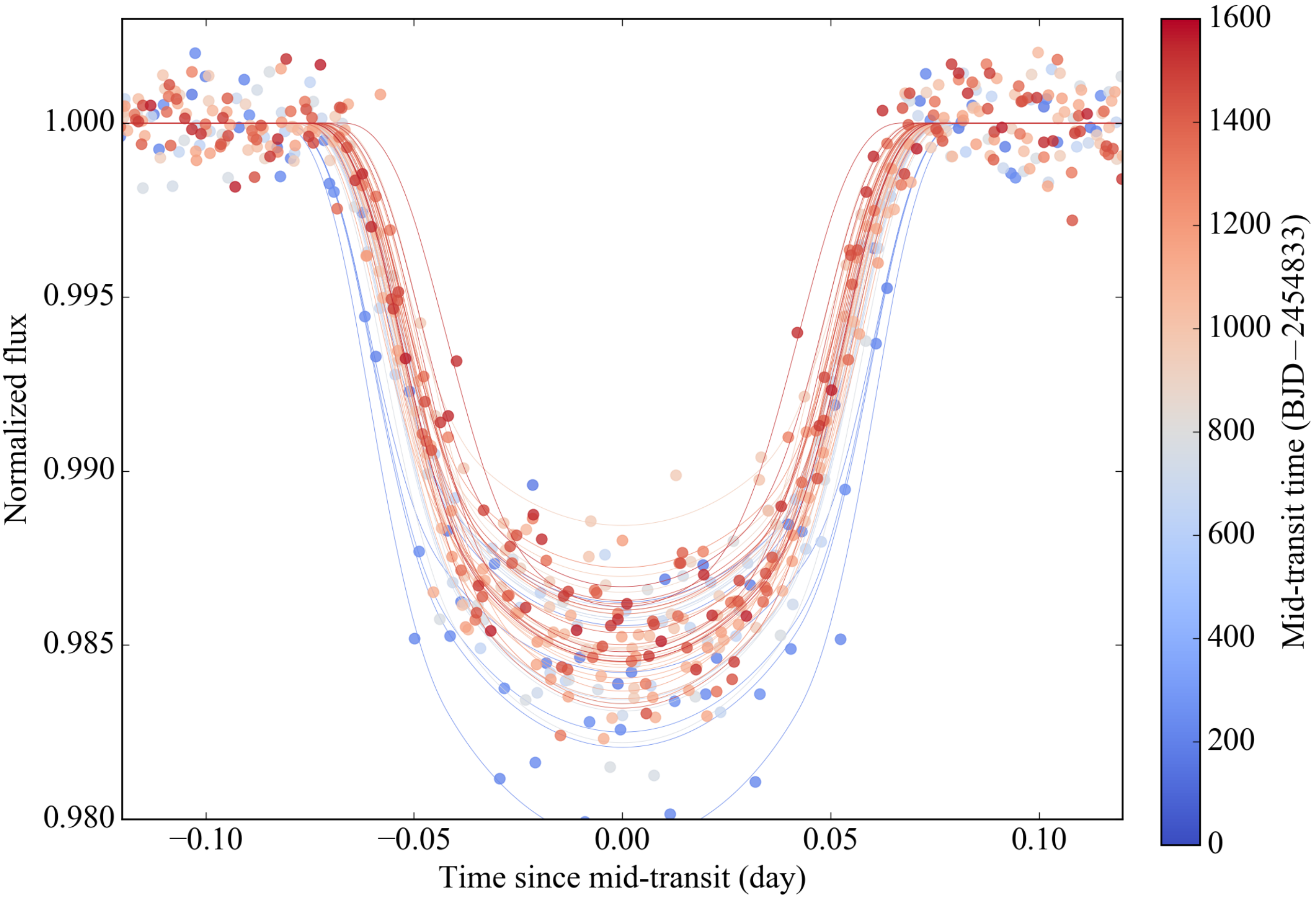}{0.49\textwidth}{(b) Kepler-693b}
	\caption{Detrended and normalized transits stacked around each center for (a) Kepler-448b (short-cadence data) and (b) Kepler-693b (long-cadence data). Filled circles show the data points, and the solid lines (only in panel b) show the best-fit transit models. The colors correspond to the mid-transit times, as shown in the right bar. Color gradations around the ingress and egress in panel (b) illustrate the systematic decrease in the transit duration (middle panel in the right column of Figure \ref{fig:tc_dur_rp}).
	}
	\label{fig:transits_koi824}
\end{figure*}

For the dynamical modeling taking into account the possible orbital misalignment, I reanalyze 
the transit light curves of Kepler-448b and Kepler-693b to derive both TTVs and TDVs (Section \ref{ssec:targets_ttv_tdv}), consistently with the other transit parameters (Tables \ref{tab:tc_dur_rp_koi12}--\ref{tab:tpars_koi824}). Here I also fit the planet-to-star radius ratio, $R_\p/R_\star$, so that the 
possible transit depth modulation due to the different crowding depending on observing seasons
does not mimic duration variations \citep[\textit{cf.}][]{2017AJ....153...45M}.
As shown in Figure \ref{fig:tc_dur_rp}, I find significant TTVs consisting of a long-term modulation and a short-term, sharp feature (top panels). For Kepler-448b, the spike-like feature is more clear than that reported in \citet{2016ApJS..225....9H}, presumably owing to a more careful treatment of the local baseline modulation. No significant correlation is found between TTVs and the local light-curve slope or the fitted $R_{\rm p}/R_\star$, which supports the physical origin rather than due to star spots \citep{2015ApJ...800..142M}. I show in Sections \ref{sec:koi12} and \ref{sec:koi824} that this unusual feature is reproduced by a periastron passage of an outer non-transiting companion in an eccentric orbit. I also identify a significant ($\sim5\sigma$) linear trend in the transit duration of Kepler-693b, which points to a mutual orbit misalignment. The duration change is also clear in Figure \ref{fig:transits_koi824}, where each detrended and normalized transit is overplotted around its center, along with the best-fit model.

\subsection{Iterative Determination of Transit Times}\label{ssec:targets_ttv}

In the systematic TTV search for WJs, I analyze Pre-search Data Conditioning Simple Aperture Photometry (PDCSAP) fluxes obtained with the long-cadence (LC) mode, retrieved from the NASA exoplanet archive. I follow the iterative method as described by \citet{2015ApJ...805...28M} to derive the consistent transit times and light-curve shapes. The method consists of (1) the determination of the central time for each transit and (2) the refinement of the transit parameters by fitting the mean transit light curve. The fitting in this subsection is performed by minimizing the standard $\chi^2$ \citep{2009ASPC..411..251M} defined as the squared sum of the difference between the model and data divided by the PDCSAP-flux error.

In the first step, I fit the data segments of three times the total duration centered at each transit. I adopt the \citet{2002ApJ...580L.171M} model for the quadratic limb-darkening law generated with {\tt pyTransit} package \citep{2015MNRAS.450.3233P}, multiplied by a second-order polynomial function of time to take into account the local baseline modulation. The fitting is repeated iteratively removing $3\sigma$ outliers. Here a circular orbit is assumed and the values of central transit time $t_{\rm c}$ and three coefficients of the polynomial are fitted, while the other parameters (two limb-darkening coefficients $q_1$ and $q_2$ defined in \citet{2013MNRAS.435.2152K}, mean stellar density $\rho_\star$, transit impact parameter $b$, planet-to-star radius ratio $R_\p/R_\star$, and orbital period $P$) are fixed. 

In the second step, I shift each transit by the value of $t_{\rm c}$ to align all the transits around time zero. Then the data are averaged into bins of $0.05$ times the transit duration \red{(to reduce the computational time)}, where the value and error of each bin is set to be the median and $1.4826$ times median absolute deviation of the flux values in the bin. The resulting ``mean" light curve is again fitted with the same \citet{2002ApJ...580L.171M} model for its central time $t_0$, normalization constant, $q_1$, $q_2$, $R_{\rm p}/R_\star$, $\rho_\star$, and $b$ assuming a circular orbit and fixing the period at the value refined in step (1); only in the first iteration, I adopt a quadratic function of time as a baseline and fit three coefficients rather than one normalization. 
The 
values of $q_1$, $q_2$, $R_{\rm p}/R_\star$, $\rho_\star$, and $b$ obtained from this fitting is used for the first step of the next iteration.

I perform five iterations for each KOI, starting with the second step based on the transit light curve phase-folded at the period given in the KOI catalog \citep{2016ApJS..224...12C}. I fit the resulting transit times with straight lines to search for any TTVs and identify the candidates mentioned above.

\subsection{Transit Times and Durations of Kepler-448b and Kepler-693b}\label{ssec:targets_ttv_tdv}

For Kepler-448b and Kepler-693b, I perform more intensive analyses taking into account possible variations of transit durations using the short-cadence (SC) data if available. For Kepler-448, I use the SC data for all the quarters, while for Kepler-693 I combine both the LC (Q$2$, Q$7$, Q$8$, Q$10$--Q$12$, Q$14$--$16$) and SC (Q$14$--Q$16$) data.

The method is the same as in Section \ref{ssec:targets_ttv} except for the following differences. I additionally fit $R_\p/R_\star$ and $b$ in the first step and repeated $10$ iterations. Here I fit $R_\p/R_\star$ so that the seasonal variation in the transit depths not be misinterpreted as the duration variation \citep[e.g.,][]{2013ApJ...774L..19V, 2015ApJ...805...28M}. The resulting $b$ and $R_{\rm p}/R_\star$ are combined with $\rho_\star$ and $P$ to yield the transit duration $T$ as the average of the total and full durations \citep[see equations 14 and 15 of][]{2011exop.book...55W}. The errors in $t_{\rm c}$ and $T$ are scaled by the square root of the reduced chi-squared $\chi^2_\nu$ of the best-fit model to enforce $\chi^2_\nu=1$. When the mean light curve is fitted, the $\chi^2$ minimization is complemented by a short Markov Chain Monte Carlo (MCMC) chain \citep{2013PASP..125..306F},
and the data are binned into one-minute bins for the SC data. 

I also tuned the parameters specifically to each system as follows:\\
{\it Kepler-448b} --- I fit the data of $1.6$ times duration for each transit. If more than $10\%$ of data is missing in the segment, the transit is excluded from the analysis. I use fourth-order polynomial because the light curve shows short-term wiggles likely due to the stellar rotation \citep[$1.245\pm0.124\,\days$;][]{2013ApJ...775L..11M}. \\
{\it Kepler-693b} --- I fit the data of twice the duration for each transit and use second-order polynomial because the light curve shows smaller variability than Kepler-448. I omit the transits with more than $10\%$ gaps for the SC data, while for the LC data I omitted those with more than $30\%$. The SC and LC data are analyzed independently and the resulting transit times and durations are averaged to give one measurement if both are available. For each data, I compute $\sqrt{\sigma_{\rm LC}^2+\sigma_{\rm SC}^2}/2$ and half of the difference between the LC and SC values, and assign the larger of the two as its error. \red{The former was the case for most of the points, while the latter process helped mitigate the effect of a few outliers caused presumably by local features in the light curve.}

The resulting transit times, durations, and radius ratios are summarized in Figure \ref{fig:tc_dur_rp} and Tables \ref{tab:tc_dur_rp_koi12} and \ref{tab:tc_dur_rp_koi824}. I also fit the mean transit light curve from the final iteration (binned version of those in Figure \ref{fig:transits_koi824}) with an MCMC including an additional Gaussian error in quadrature (denoted as ``photometric jitter") and derive the posteriors for the transit parameters summarized in Tables \ref{tab:tpars_koi12} and \ref{tab:tpars_koi824}. For Kepler-448b, the result agreed well with those from the least-square fit with its standard errors scaled by the square root of $\chi^2_\nu$. For Kepler-693b, I obtain skewed posteriors and larger errors; the parameters are better determined by the LC data that include more transits.

\begin{deluxetable*}{lcccc}
	\tablewidth{0pt}
	\tablecolumns{2}
	\tablecaption{Transit Times, Durations, and Radius Ratios of Kepler-448b\label{tab:tc_dur_rp_koi12}}
	\tablehead{
		\colhead{Transit number}  
		& \colhead{Transit time ($\kbjd$)} & \colhead{Transit duration (day)}
		& \colhead{Planet-to-star radius ratio}
	}
	\startdata
	$-1$	&	$128.7420\pm0.0001$	&	$0.2808\pm0.0003$	&	$0.0920\pm0.0004$\\
	\nodata\\
	\enddata
	\tablecomments{Quoted uncertainties are the standard errors derived from the covariance
	matrix scaled by $\sqrt{\chi^2/\mathrm{d.o.f.}}$ of the fit. This table is published in its entirety in the machine-readable format. A portion is shown here for guidance regarding its form and content.}
\end{deluxetable*}

\begin{deluxetable*}{lcccc}
	\tablewidth{0pt}
	\tablecolumns{2}
	\tablecaption{Transit Times, Durations, and Radius Ratios of Kepler-693b \label{tab:tc_dur_rp_koi824}}
	\tablehead{
		\colhead{Transit number}  
		& \colhead{Transit time ($\kbjd$)} & \colhead{Transit duration (day)}
		& \colhead{Planet-to-star radius ratio}
	}
	\startdata
	$0$	&	$173.609\pm0.002$	&	$0.115\pm0.007$	&	$0.115\pm0.008$\\
	\nodata\\
	\enddata
	\tablecomments{Quoted uncertainties are the standard errors derived from the covariance matrix scaled by $\sqrt{\chi^2/\mathrm{d.o.f.}}$ of the fit, or are based on the combination of the LC and SC data (see Section \ref{ssec:targets_ttv_tdv}). This table is published in its entirety in the machine-readable format. A portion is shown here for guidance regarding its form and content.}
\end{deluxetable*}

\begin{deluxetable}{lc}
	\tablewidth{0pt}
	\tablecolumns{2}
	\tablecaption{Transit Parameters of Kepler-448b Based on Short-cadence Data\label{tab:tpars_koi12}}
	\tablehead{
		\colhead{Parameter} & \colhead{Value}
	}
	\startdata
	\multicolumn{2}{l}{({\it Parameters from the Mean Transit Light Curve})\tablenotemark{a}}\\
	$q_1=(u_1+u_2)^2$	&	$0.199(7)$\\
	$q_2=\frac{u_1}{2(u_1+u_2)}$	&	$0.39(2)$\\
	Center of the mean transit (day)	&	$-0.00001(2)$\\
	$R_\p/R_\star$	&	$0.09044(7)$\\
	Normalization	&	$0.999999(3)$\\
	Transit impact parameter & $0.373(6)$\\
	Mean stellar density ($\gcc$) & $0.393(3)$\\
	Photometric jitter & $0.000036(2)$\\
	$(a/R_\star)_{e=0}$	&	$18.77(4)$\\
	$u_1$	&	$0.348(9)$\\
	$u_2$	&	$0.10(2)$\\
	\multicolumn{2}{l}{({\it Mean Orbital Period and Transit Epoch})\tablenotemark{b}}\\
	$t_0$ ($\mathrm{BJD_{TDB}}$)	&	$2454979.5961(2)$\\
	$P$ (day)		&	$17.855234(4)$\\
	\enddata
	\tablenotetext{a}{Median and $68\%$ credible interval of the MCMC posteriors
	from the mean transit light curve. 
	Here	the circular orbit is assumed to relate mean stellar density 
	to the semi-major axis over stellar radius, $a/R_\star$. 
	The limb-darkening coefficients $u_1$ and $u_2$ are converted from $q_1$ and $q_2$.}
	\tablenotetext{b}{Obtained by linearly fitting the series of transit times in Table \ref{tab:tc_dur_rp_koi12}. Errors are scaled by $\sqrt{\chi^2/\mathrm{d.o.f.}}$.}
	\tablecomments{Parentheses after values denote uncertainties in the last digit.}
\end{deluxetable}

\begin{deluxetable}{lc}
	\tablewidth{0pt}
	\tablecolumns{2}
	\tablecaption{Transit Parameters of Kepler-693b Based on Long-cadence Data\label{tab:tpars_koi824}}
	\tablehead{
		\colhead{Parameter} & \colhead{Value}
	}
	\startdata
	\multicolumn{2}{l}{({\it Parameters from the Mean Transit Light Curve})\tablenotemark{a}}\\
	$q_1=(u_1+u_2)^2$	&	$0.6^{+0.3}_{-0.2}$\\
	$q_2=\frac{u_1}{2(u_1+u_2)}$	&	$0.4^{+0.3}_{-0.2}$\\
	Center of the mean transit (day)	&	$0.0000(3)$\\
	$R_\p/R_\star$	&	$0.117^{(+4)}_{(-6)}$\\
	Normalization	&	$1.00000(8)$\\
	Transit impact parameter & $0.63^{+0.06}_{-0.15}$\\
	Mean stellar density ($\gcc$) & $3.0^{+1.0}_{-0.5}$\\
	Photometric jitter & $0.00074(7)$\\
	$(a/R_\star)_{e=0}$	&	$34^{+4}_{-2}$\\
	$u_1$	&	$0.7^{+0.2}_{-0.3}$\\
	$u_2$	&	$0.1^{+0.3}_{-0.4}$\\
	\multicolumn{2}{l}{({\it Mean Orbital Period and Transit Epoch})\tablenotemark{b}}\\
	$t_0$ ($\mathrm{BJD_{TDB}}$)	&	$2455006.613(1)$\\
	$P$ (day)		&	$15.37566(3)$\\
	\enddata
	\tablenotetext{a}{Median and $68\%$ credible interval of the MCMC posteriors
	from the mean transit light curve. 
	Here	the circular orbit is assumed to relate mean stellar density 
	to the semi-major axis over stellar radius, $a/R_\star$. 
	The limb-darkening coefficients $u_1$ and $u_2$ are converted from $q_1$ and $q_2$.}
	\tablenotetext{b}{Obtained by linearly fitting the series of transit times in Table \ref{tab:tc_dur_rp_koi824}. Errors are scaled by $\sqrt{\chi^2/\mathrm{d.o.f.}}$.}
	\tablecomments{Parentheses after values denote uncertainties in the last digit.}
\end{deluxetable}

\section{Dynamical Modeling of TTVs and TDVs: Method}\label{sec:method}

\subsection{Model Assumptions\label{ssec:method_ass}}

The TTVs observed in the two systems (Figure \ref{fig:tc_dur_rp}) are clearly different from the sinusoidal signal due to a companion near a mean-motion resonance \citep{2012ApJ...761..122L}. 
In particular, rapid timing variations on a short timescale suggest that the perturbing companions' orbits are eccentric. In addition, such a feature is observed only once for each system, and so
the companions must be far outside the WJs' orbits. 
Thus I assume a hierarchical three-body system as the only viable configuration and model the observed TTVs and TDVs to derive the outer companions' masses and orbits.

I only consider the Newtonian gravity between the three bodies regarded as point masses (see Section \ref{sssec:method_joint_tdv} for justification), as well as the finite light-travel time in computing timings. To better take into account the hierarchy of the system, I adopt the Jacobi coordinates to describe their orbits: the inner orbit (denoted by the subscript $1$) is defined by the relative motion of the inner planet around the host star, while the outer one (denoted by the subscript $2$) is the motion of the outer companion relative to the center of mass of the inner two bodies. The sky plane is chosen to be the reference plane, to which arguments of periastron and the line of nodes are referred. The direction of $+Z$-axis (which matters the definition of ``ascending" node) is taken toward the observer.

\subsection{Bayesian Framework\label{ssec:method_bayes}}

I derive the posterior probability density function (PDF) for the set of system parameters $\bm{\theta}$ 
conditioned on the observed data $\bm{d}$: 
\begin{equation}
	\label{eq:bayes}
	p(\bm{\theta}|\bm{d}, I) 
	= \frac{1}{\mathcal{Z}}\,p(\bm{d}|\bm{\theta}, I)\,p(\bm{\theta}|I),
\end{equation}
where $I$ denotes the prior information. The normalization factor 
\begin{equation}
	\label{eq:evidence}
	\mathcal{Z} \equiv \int p(\bm{d}|\bm{\theta}, I)\,p(\bm{\theta}|I)\mathrm{d}\bm{\theta}
\end{equation}
is called the global likelihood or evidence, which represents the plausibility of the model. The prior PDF $p(\bm{\theta}|I)$ is given as a product of the prior PDFs for each parameter, which are assumed to be independent; they are presented in Tables \ref{tab:photod_koi12} and \ref{tab:photod_koi824}. The likelihood $\mathcal{L}\equiv p(\bm{d}|\bm{\theta}, I)$ is defined and computed as described in Sections \ref{ssec:method_ttv} and \ref{ssec:method_joint}. 

To invert the observed signals, I utilize the nested-sampling algorithm {\tt MultiNest} \citep{2009MNRAS.398.1601F, 2008MNRAS.384..449F, 2013arXiv1306.2144F} and its python interface {\tt PyMultiNest} \citep{2014A&A...564A.125B}, which allows us to sample the whole prior volume to identify multiple modes if any. I typically utilize $4800$ live points and default sampling efficiency of $0.8$, keep updating the live points until the evidence tolerance of $0.5$ is achieved, and allow for the detection of multiple posterior modes. When multiple modes are detected, {\tt MultiNest} also computes evidence corresponding to each mode, which is referred to as ``local" evidence.

\subsection{Procedure for Finding Solutions\label{ssec:method_procedure}}

To check on the reliability of our numerical scheme and to find all the possible posterior modes, I adopt the following procedure. First, I use the analytic TTV formula for a hierarchical triple system \citep[][Appendix \ref{app:formula}]{2015MNRAS.448..946B} to fit only the TTV signal. This allows us to search a wide region of the parameter space and resulted in one global mode containing two peaks. The resulting solution was also found to be consistent with a rough analytical estimate based on the observed TTV features (see, e.g., Section \ref{ssec:koi12_ttv}). Then, I fit the same TTVs numerically adopting a slightly narrower prior range that well incorporates the global mode found from the analytic fit. The resulting posterior was found to be consistent with the one derived from the analytic fit. This agreement validates the numerical scheme I rely on. Finally, the same numerical method is used to model both TTVs and TDVs simultaneously to determine the physical and geometric properties of the outer companions. In Sections \ref{sec:koi12} and \ref{sec:koi824}, I mainly report and discuss the results from the final fit, while the analytic and numerical posteriors from TTVs alone are found in Appendix \ref{app:ttv} for comparison.

\subsection{TTV Modeling\label{ssec:method_ttv}}

The likelihood for the TTV modeling is defined as follows:
\begin{align}
	\notag
	\mathcal{L}_{\rm TTV}\equiv
	\prod_i \frac{1}{\sqrt{2\pi(\sigma_{t,i}^2 + \sigma_{\rm TTV}^2)}}
	\,\exp \left[ -\frac{(t_i - t_i^\mathrm{model})^2}{2(\sigma_{t,i}^2 + \sigma_{\rm TTV}^2)}\right],
\end{align}
where $t_i$ and $\sigma_{t,i}$ are the transit times and their errors obtained in Section \ref{ssec:targets_ttv_tdv} (Tables \ref{tab:tc_dur_rp_koi12} and \ref{tab:tc_dur_rp_koi824}). I also include $\sigma_{\rm TTV}$ as a model parameter that takes into account the additional scatter and marginalize over it to obtain more conservative constraint; it turns out that this parameter is not correlated with any other physical parameters (Figures \ref{fig:corner_koi12} and \ref{fig:corner_koi824}) and hence does not affect the global shape of the joint PDF. 

The model transit times $t^{\rm model}_i$ are computed both analytically and numerically as mentioned in Section \ref{ssec:method_procedure}; this is for cross-validation, as well as to obtain insights into how the parameters are determined. In both cases, the model includes $14$ parameters in addition to $\sigma_{\rm TTV}$ (see Tables \ref{tab:ttv_koi12} and \ref{tab:ttv_koi824}): orbital period, orbital phase, eccentricity, and argument of periastron for both inner and outer orbits; cosine of the outer orbit's inclination (inner one is fixed to be $0$ \red{for simplicity; this does not affect the result}); difference in the longitudes of the ascending node; and masses of the host star and outer companion (here I fix the inner planet's mass to be $1M_{\rm Jup}$ \red{as it is not determined from TTVs}). While I use the time of inferior conjunction $t_{\rm ic}$ and orbital period $P$ for the inner orbit, I choose the time of periastron passage $\tau_2$ and periastron distance relative to the inner semi-major axis $q_2/a_1$ to specify the outer orbital phase and period. This is because the latter two parameters are more directly related to the position and duration of the ``spike" in the observed TTVs than $t_{\rm ic}$ and $P$, and thus better determined. I fit the mass scale of the whole system as well because I include the light-travel time effect; in practice, however, the observed TTVs are insensitive to this parameter and its value is solely constrained by the prior knowledge.

I first compute analytic transit times using the formula by \citet{2011A&A...528A..53B, 2015MNRAS.448..946B} developed for hierarchical planetary systems and triple-star systems (Appendix \ref{app:formula}). I include the light-travel time effect (LTTE) and $P_2$-timescale dynamical effect due to the quadrupole component of the perturbing potential; I did not find notable changes even with the octupole terms. The former is due to the motion of the inner binary (here the central star orbited by the inner planet), which changes finite time for the light emitted from the star and blocked by the planet to travel to us. The latter process involves the actual variation in the orbital period of the inner binary due to the tidal gravitational field exerted by the companion. Both of these effects are routinely observed in triple-star systems containing eclipsing binaries \citep[e.g.,][]{2013ApJ...768...33R, 2015ApJ...806L..37M}.

Numerical TTVs are computed using {\tt TTVFast} code \citep{2014ApJ...787..132D} by integrating the orbits of the three bodies and finding the times at which sky-projected distance between the star and the inner planet, $d_{\rm sky}$, becomes minimum. I choose the time step to be $0.3\,\mathrm{days}$ for Kepler-448 and $0.1\,\mathrm{days}$ for Kepler-693, which are roughly $1/60$ and $1/150$ of the inner orbital periods, respectively. To compute transit times, I modify the default output of the {\tt TTVFast} code to take into account the effect of light-travel time using the line-of-sight coordinate of the center of mass of the inner binary (i.e., central star and inner planet).

\subsection{Joint TTV and TDV Modeling\label{ssec:method_joint}} 

The joint TTV and TDV modeling was performed using the following likelihood:
\begin{align}
	\notag
	\mathcal{L}=&\mathcal{L}_{\rm TTV}\\
	\notag
	&\times
	\prod_j \frac{1}{\sqrt{2\pi(\sigma_{T,j}^2 + \sigma_{\rm dur}^2)}}
	\,\exp \left[ -\frac{(T_j - T_j^\mathrm{model})^2}{2(\sigma_{T,j}^2 + \sigma_{\rm dur}^2)}\right]\\
	\notag
	&\times
	\frac{1}{\sqrt{2\pi\sigma_b^2}}
	\,\exp \left[ -\frac{(b_{\rm mean} - b_{\rm mean}^\mathrm{model})^2}{2\sigma_b^2}\right],
\end{align}
where the second row is defined analogously to $\mathcal{L}_{\rm TTV}$ with the transit time $t$ replaced by the transit duration $T$. I also include the constraint on the transit impact parameter $b_{\rm mean}$ from the mean transit light curve.\footnote{\red{Here we can use $b_{\rm mean}$ derived assuming a circular orbit because the parameter is based on the shape of the light curve alone. A possible bias due to the non-zero eccentricity is absorbed in the fitted stellar density, which is not used in any of the following analyses. See, e.g., Eqn. (28) and (29) of \citet{2011exop.book...55W}.}} The TDV modeling additionally requires the mean density of the host star $\rho_\star$ (equivalently stellar radius) and the ``jitter" for durations $\sigma_{\rm dur}$, and $\bm{\theta}$ consists of $17$ parameters listed in Tables \ref{tab:photod_koi12} and \ref{tab:photod_koi824} as fitted parameters. Here I fit inner orbit's inclination and inner planet's mass, although the latter is not constrained at all by the data.

This joint modeling is performed only numerically using {\tt TTVFast}. In addition to the transit times above, I use another default output, $d_{\rm sky}$ at each transit center, to obtain the model transit impact parameter $b=d_{\rm sky}/R_\star$, where the stellar radius $R_\star$ is given by $(3m_\star/4\pi\rho_\star)^{1/3}$; here I adopt $b$ at the transit around the center of the observing duration as $b^{\rm model}_{\rm mean}$. The transit durations are obtained from yet another default output of the code $v_{\rm sky}$, the sky-projected relative star--planet velocity at the transit center, via $T=2\sqrt{1-b^2}R_\star/v_{\rm sky}$.

\subsubsection{Additional TDV Sources\label{sssec:method_joint_tdv}} 

While I only consider the Newtonian gravitational interaction between point masses, transit durations may also be affected by secular orbit variation due to stellar quadrupole moment and/or general relativistic effect. Indeed, we expect that Kepler-448 has a relatively large quadrupole moment $J_2$ due to its rapid rotation \citep{2013ApJ...775L..11M, 2015A&A...579A..55B}, and the inner orbits are moderately eccentric in both systems as will be shown later. Here we show that those effects on transit durations are negligible \citep[the effects on transit times are even smaller; see][]{2002ApJ...564.1019M}.

The duration drift due to the nodal precession is given by \citep{2002ApJ...564.1019M}
\begin{align}
	|\dot T| \leq 3J_2 \left(\frac{a}{R_\star}\right)^{-2} \frac{b}{\sqrt{1-b^2}}\,|\sin \lambda|.
\end{align}
where $\lambda$ is the sky-projected stellar obliquity. For Kepler-448b with $b\simeq0.4$, $a/R_\star\simeq20$, and $\lambda\simeq10\degr$, this results in the duration change over four years 
\begin{align}
	\Delta T\lesssim 2\times10^{-3}\,\mathrm{hr} \left(\frac{J_2}{10^{-4}}\right),
\end{align}
\red{where the quoted value of $J_2$ is motivated by theoretical modeling and observational inference for a star with similar parameters \citep{2012MNRAS.421L.122S, 2015ApJ...805...28M}. This $\Delta T$ is smaller than the measurement precision.} The same is also true for Kepler-693b, for which $J_2$ is likely smaller and measurements of $T$ are less precise. 

The drift caused by general relativistic apsidal precession is \citep{2002ApJ...564.1019M}
\begin{align}
	\dot T \leq 4e \left(\frac{a}{R_\star}\right)^{-1} \sqrt{1-b^2} \times \frac{3}{1-e^2}\left(\frac{na}{c}\right)^2
\end{align}
or
\begin{equation}
	\Delta T\lesssim2\times10^{-3}\,\mathrm{hr}\left(\frac{e}{1-e^2}\right)
\end{equation}
over four years; this is also negligible. Note that possible apsidal precession induced by the gravitational perturbation from the outer companion is already taken into account in our Newtonian model.

\section{Results: Kepler-448/KOI-12}\label{sec:koi12}

\begin{figure*}
	\centering
	\includegraphics[width=0.9\textwidth]{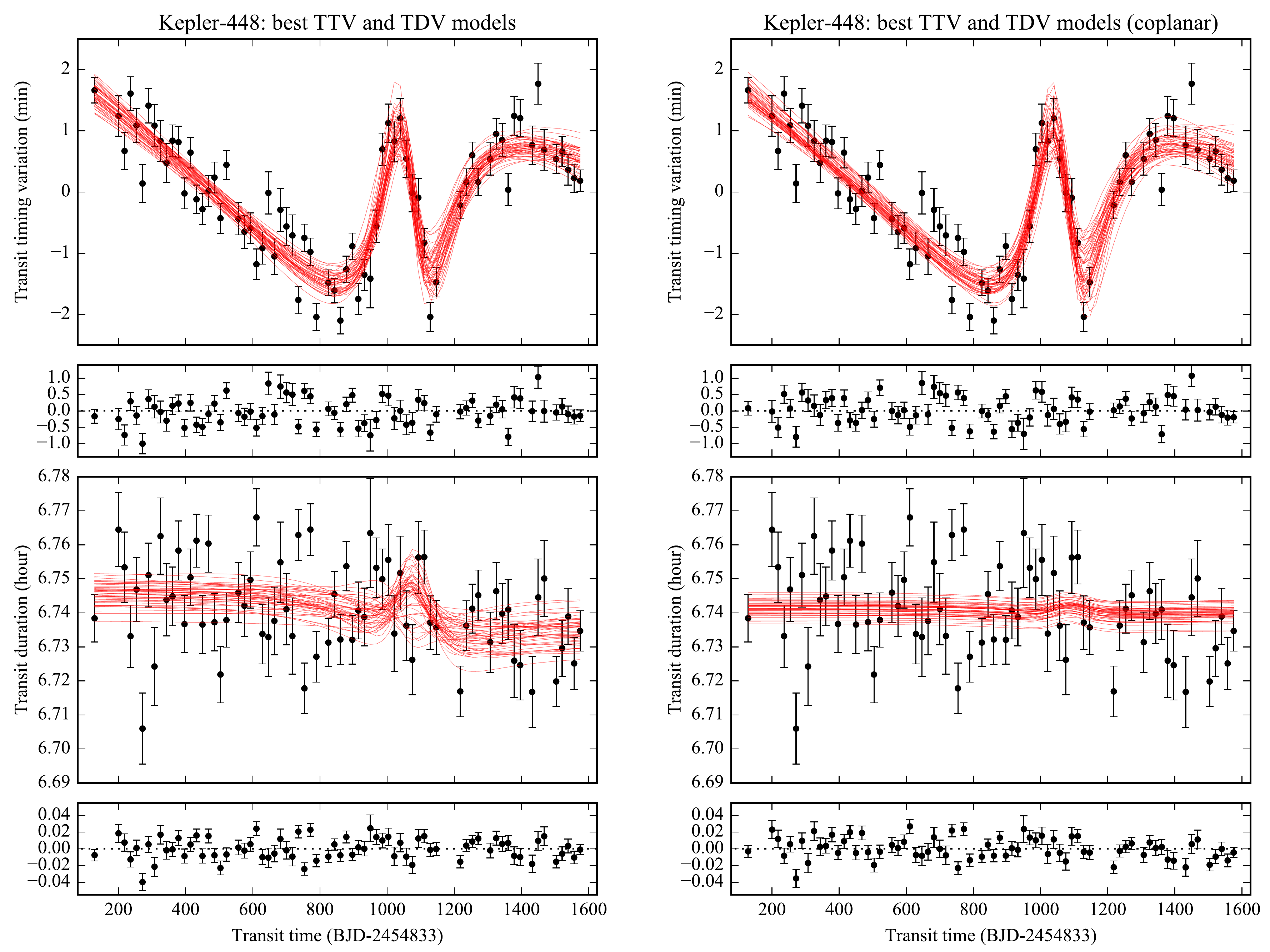}
	\caption{Dynamical models of the observed TTVs and TDVs of Kepler-448b.
	Red solid lines show $50$ models randomly sampled from
	the joint posterior distribution obtained by fitting the data
	(black circles with error bars).
	The left column corresponds to the model which allows the mutual inclination to vary,
	while the right column shows the result when it is fixed to be zero.
	}
	\label{fig:bestfit_koi12}
\end{figure*}

The resulting posterior PDF from the {\tt MultiNest} analysis and the corresponding models are shown in Figures \ref{fig:bestfit_koi12} (red solid lines) and Figure \ref{fig:corner_koi12}. The summary statistics (median and $68\%$ credible interval) of the marginal posteriors as well as the priors adopted for those parameters are given in Table \ref{tab:photod_koi12}. The solution consists of two separated posterior modes (denoted as Solution 1 and Solution 2) that reproduce the observed TTVs and TDVs equally well, without any significant difference in the local evidence computed for each mode.

In both solutions, the outer companion (we tentatively call it Kepler-448c) is likely to have a mass in the brown-dwarf regime ($22^{+7}_{-5}\,M_{\rm Jup}$) and reside in a highly eccentric orbit close to the inner WJ. In spite of the small outer orbit, its pericenter distance $1.46^{+0.07}_{-0.06}\,\mathrm{au}$ is more than nine times larger than the inner semi-major axis and well within the stable regime \citep{2001MNRAS.321..398M}. The system has one of the smallest binary separations compared to the planet apastron $0.21^{+0.05}_{-0.04}\,\mathrm{au}$ among the known planetary systems with (sub-)stellar companions; see figure 8 of \citet{2017MNRAS.tmp..167T}.

I also find a modest eccentricity $0.34^{+0.08}_{-0.07}$ for the inner WJ from the combination of timing and duration information. While the TTVs alone favor an even larger value (Appendix \ref{app:ttv}), transit duration combined with the spectroscopic prior on $\rho_\star$ lowers it. The inner WJ mass is also floated between $0.1\mathchar`-10\,M_{\rm Jup}$ but no constraint better than that from the RV data \citep[$<10\,M_{\rm Jup}$ as the $3\sigma$ upper bound;][]{2015A&A...579A..55B} is found. This is because the TTVs of the inner WJ does not depend on its own mass to the first order.

\begin{figure}
	\centering
	\includegraphics[width=\linewidth]{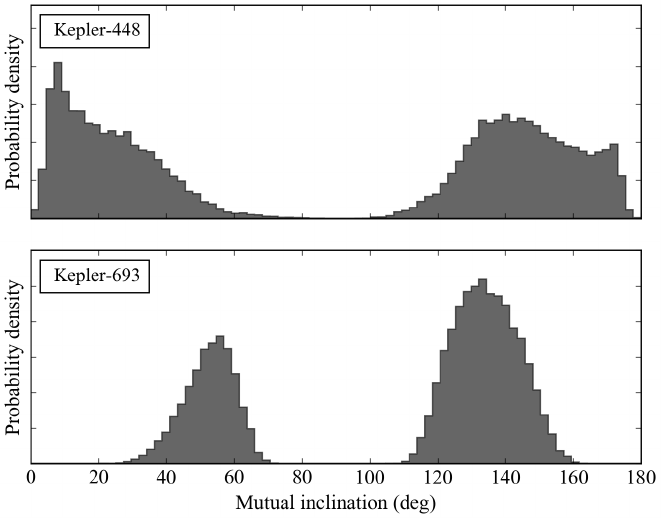}
	\caption{Marginal posterior PDFs for the mutual inclination between the inner and outer orbits from the dynamical TTV and TDV modeling (Sections \ref{sec:koi12} and \ref{sec:koi824}). {\it Top} --- Kepler-448. {\it Bottom} --- Kepler-693.
	}
	\label{fig:imut}
\end{figure}

While the above constraints essentially come from TTVs, TDVs possibly allow us to constrain the mutual inclination $i_{\rm 21}$ between the inner and outer orbits. In fact, the two solutions are different in this respect:  Solution 1 is ``prograde", i.e., the two orbits are in the same directions, while Solution 2 corresponds to a ``retrograde" configuration. Non-zero mutual inclinations are slightly favored in both solutions with the posterior probability that $39\fdg2<i_{21}<140\fdg8$ being $27\%$. However, it is more or less due to the large prior volume corresponding to the misaligned solution and a wide range of the mutual inclination is allowed by the data (Figure \ref{fig:imut}). In particular, the data are totally consistent with the coplanar configuration, as visually illustrated in the right panels of Figure \ref{fig:bestfit_koi12}. I found the evidence for the coplanar model is not significantly different from the misaligned case. Only the configuration with $i_{\rm 21}\simeq90\degr$ is slightly disfavored due to the lack of a large TDV expected from the strong perturbation in the direction perpendicular to the orbital plane.

The joint TTV/TDV analysis also allows for constraining the impact parameter during a possible occultation, $b_{\rm occ}$,  based on that during the transit as well as the eccentricity and argument of periastron of the inner orbit. I find $b_{\rm occ}=0.22\pm0.02$, which means that the secondary eclipse should have been detected if Kepler-448b was a star. The fact strengthens the planetary interpretation by \citet{2015A&A...579A..55B} who confirmed $m_\mathrm{b}<10\,M_{\rm Jup}$ at the $3\sigma$ level.

The following subsections detail the prior information and interpretation of TTVs and TDVs.

\begin{deluxetable*}{lcccc}
	\tablewidth{0pt}
	\tablecolumns{5}
	\tablecaption{Parameters of the Kepler-448 System from the Dynamical TTV and TDV Analysis\label{tab:photod_koi12}}
	\tablehead{
		\colhead{Parameter} & \colhead{Solution 1} & \colhead{Solution 2} & \colhead{Combined} &  \colhead{Prior}
	}
	\startdata
	{\bf Fitted Parameters}\\
	({\it Inner Orbit})\\
	1. Time of inferior conjunction 
		& $128.7418^{(+2)}_{(-3)}$ 
		& $128.7413^{(+2)}_{(-2)}$
		& $128.7415^{(+4)}_{(-3)}$
		& $\uni(128.73, 128.75)$\\
		\quad $t_{\rm ic, 1}$ ($\kbjd$) & & &\\
	2. Orbital period $P_1$ (day)
		& $17.855183^{(+9)}_{(-6)}$ 
		& $17.855181^{(+7)}_{(-5)}$
		& $17.855182^{(+8)}_{(-6)}$
		& $\luni(17.8551, 17.8553)$\\
	3. Orbital eccentricity $e_1$
		& $0.34^{+0.07}_{-0.06}$ 
		& $0.35^{+0.09}_{-0.08}$
		& $0.34^{+0.08}_{-0.07}$
		& $\uni(0, 0.7)$\\
	4. Argument of periastron $\omega_1$ (deg)
		& $-49^{+9}_{-21}$
		& $-131^{+18}_{-9}$ 
		& \nodata
		& $\uni(-180, 180)$\\
	5. Cosine of orbital inclination $\cos i_1$\tablenotemark{a}
		& $0.013(1)$ 
		& $0.013(1)$
		& $0.013(1)$
		& $\uni(0, 0.02)$\tablenotemark{a}\\
	({\it Outer Orbit})\\
	6. Time of the periastron passage
		& $1076^{+10}_{-11}$ 
		& $1080^{+10}_{-11}$ 
		& $1078^{+10}_{-11}$
		& $\uni(900, 1200)$\\
		\quad $\tau_2$ ($\kbjd$) &&&\\
	7. Periastron distance over 
		& $9.5^{+0.4}_{-0.3}$ 
		& $9.4^{+0.4}_{-0.3}$ 
		& $9.4^{+0.4}_{-0.3}$
		& $\luni(5, 20)$\\
		\quad inner semi-major axis $q_2/a_1$&&&\\
	8. Orbital eccentricity $e_2$
		& $0.61^{+0.11}_{-0.07}$ 
		& $0.69^{+0.12}_{-0.10}$ 
		& $0.65^{+0.13}_{-0.09}$
		& $\uni(0, 0.95)$\\
	9. Argument of periastron $\omega_2$ (deg)
		& $-89^{+6}_{-7}$ 
		& $-89^{+5}_{-6}$ & $-89^{+5}_{-6}$
		& $\uni(-180, 180)$\\
	10. Cosine of orbital inclination $\cos i_2$\tablenotemark{a}
		& $-0.3^{+0.3}_{-0.3}$ 
		& $-0.6^{+0.3}_{-0.2}$ & $-0.5^{+0.4}_{-0.3}$
		& $\uni(-1, 1)$\\
	11. Relative longitude of 
		& $3^{+4}_{-4}$	& $-178^{+5}_{-4}$ & \nodata
		& $\uni(-180, 180)$\\
		\quad ascending node $\Omega_{21}$ (deg)\tablenotemark{a,b} &&&\\
	({\it Physical Properties})\\
	12. Mass of Kepler-448 $m_\star$ ($M_\odot$)
		& $1.5\pm0.1$	 & $1.5\pm0.1$ & $1.5\pm0.1$  
		& $\gaus(1.51, 0.09, 0.14)$\\
	13. Mean density of 
		& $0.79^{+0.09}_{-0.09}$	& $0.81^{+0.10}_{-0.09}$
		& $0.80^{+0.10}_{-0.09}$
		& $\gaus(0.76, 0.11, 0.12)$\\
		\quad Kepler-448 $\rho_\star$ ($\gcc$) &&&\\
	14. Mass of Kepler-448b $m_\mathrm{b}$ ($M_{\rm Jup}$)\tablenotemark{c}
		& $1.1^{+3.5}_{-0.8}$ 
		& $1.1^{+3.5}_{-0.8}$ & $1.1^{+3.5}_{-0.8}$
		& $\luni(0.1, 10)$\\
	15. Mass of Kepler-448c $m_\mathrm{c}$ ($M_{\rm Jup}$)
		& $21^{+6}_{-5}$	
		& $24^{+6}_{-5}$ & $22^{+7}_{-5}$
		& $\luni(0.001M_\odot, 0.1M_\odot)$\\
	({\it Jitters})\\
	16. Transit time jitter $\sigma_{\rm TTV}$ ($10^{-4}\,\mathrm{day}$)
		& $2.2\pm0.3$	& $2.2\pm0.3$	 & $2.2\pm0.3$
		& $\luni(5\times10^{-2}, 5)$\\
	17. Transit duration jitter $\sigma_{\rm dur}$ ($10^{-4}\,\mathrm{day}$)
		& $3.8\pm0.6$		& $3.7\pm0.6$		& $3.7\pm0.6$
		& $\luni(0.1, 10)$\\
	\\
	{\bf Derived Parameters}\\
	Outer orbital period $P_2$ (day)
		& $(2.2^{+1.5}_{-0.5})\times10^3$ & $(2.9^{+3.3}_{-1.0})\times10^3$
		& $(2.5^{+2.4}_{-0.7})\times10^3$
		& \nodata\\
	Inner semi-major axis $a_1$ (au)
		&  $0.154^{(+4)}_{(-3)}$ & $0.154^{(+4)}_{(-3)}$ 
		& $0.154^{(+4)}_{(-3)}$ 
		& \nodata\\
	Outer semi-major axis $a_2$ (au)
		&  $3.8^{+1.6}_{-0.6}$ 
		& $5^{+3}_{-1}$ & $4.2^{+2.4}_{-0.9}$ & \nodata\\
	Periastron distance of 
		&  $1.47^{+0.07}_{-0.06}$ 
		&  $1.45^{+0.06}_{-0.06}$ 
		&  $1.45^{+0.07}_{-0.06}$ & \nodata\\
		\quad the outer orbit $a_2(1-e_2)$ (au) &&&\\
	Mutual orbital inclination $i_{21}$ (deg)
		& $20^{+17}_{-12}$	& $146^{+19}_{-16}$	
		& \nodata & \nodata\\
	Physical radius of Kepler-448 ($R_\odot$)
		& $1.40\pm0.06$	& $1.39^{+0.07}_{-0.06}$ 
		& $1.40^{+0.07}_{-0.06}$ & \nodata\\
	Physical radius of Kepler-448b ($R_\mathrm{Jup}$)\tablenotemark{d}
		& $1.24^{+0.06}_{-0.05}$	& $1.23^{+0.06}_{-0.06}$ 
		& $1.23^{+0.06}_{-0.05}$ & \nodata\\
	Transit impact parameter of Kepler-448b
		& $0.372(6)$ & $0.371(6)$ & $0.371(6)$ & \nodata\\
	Occultation impact parameter of Kepler-448b
		& $0.22\pm0.02$ & $0.21\pm0.02$ & $0.22\pm0.02$ & \nodata\\
	\\
	Log evidence $\ln \mathcal{Z}$ from {\tt Multinest} & $-168.16\pm0.09$ & $-168.04\pm0.09$ & $-167.40\pm0.09$\\
	\enddata
	\tablecomments{The elements of the inner and outer orbits listed here
	are Jacobian osculating elements defined at the epoch 
	$\mathrm{BJD}=2454833+120$.
	The quoted values in the `Solution' columns are the median
	and $68\%$ credible interval of the marginal posteriors.
	Parentheses after values denote uncertainties in the last digit.
	The `combined' column shows the values from the marginal posterior 
	combining the two solutions; no value is shown if the combined 
	marginal posterior is multimodal. In the prior column,
	$\uni(a, b)$ and $\luni(a, b)$ denote the (log-)uniform priors between $a$ and $b$,
	$f(x)=1/(b-a)$ and $f(x)=x^{-1}/(\ln b - \ln a)$, respectively;
	$\gaus(a, b, c)$ means the asymmetric Gaussian prior 
	with the central value $a$ and lower and upper widths $b$ and $c$.}
	\tablenotetext{a}{\red{There also exists a solution with negative $\cos i_1$. The solution is statistically indistinguishable from the one reported here, except that the signs of $\cos i_2$ and $\Omega_{21}$ are flipped (and thus $i_{21}$ remains the same; see Eqn. \ref{eq:i21}). In principle, the TTVs for the solutions with $\cos i_1>0$ and $\cos i_1<0$ are not completely identical \citep[see, e.g., A15 of][]{2015MNRAS.448..946B}. However, the difference is in practice negligibly small for a transiting system because the effect is proportional to $\cot i_1$. I confirmed that this is indeed the case by performing the same numerical fit with the prior on $\cos i_1$ replaced by $\uni(-0.02, 0)$.}}
	\tablenotetext{b}{Referenced to the ascending node of the inner orbit,
	whose direction is arbitrary.}
	\tablenotetext{c}{This parameter is not constrained by the data; posterior is identical to the log-uniform prior.}
	\tablenotetext{d}{Derived from the posterior of $R_\star$ and mean and standard deviation of $R_{\rm p}/R_\star$ from individual transits (Table \ref{tab:tpars_koi12}).}
\end{deluxetable*}

\subsection{Adopted Parameters}\label{ssec:koi12_prior}

I adopt $m_\star=1.51^{+0.14}_{-0.09}\,M_\odot$ and $\rho_\star=0.758^{+0.12}_{-0.11}\,\gcc$ (converted from the radius $r_\star=1.41\pm0.06\,R_\odot$) as the priors based on the spectroscopic values by \citet{2015A&A...579A..55B}. The values agree with, and are more precise than, the latest KIC values 
\citep{2016arXiv160904128M}: $m_\star=1.386^{+0.093}_{-0.076}\,M_\odot$, $r_\star=1.367^{+0.267}_{-0.067}\,R_\odot$, and $\rho_\star=0.7647^{+0.09607}_{-0.2874}\,\gcc$. Note that I do not use the values from the joint analysis in \citet{2015A&A...579A..55B} because they are derived assuming a circular orbit for the inner transiting planet, which turns out unlikely from our dynamical modeling. I adopt $b_{\rm mean}=0.373$ and $\sigma_b=0.006$ based on the posterior from the mean transit light curve (Table \ref{tab:tpars_koi12}).

\subsection{Constraints from TTVs}\label{ssec:koi12_ttv}

The mass and some elements of the outer orbit ($\tau_2$, $q_2/a_1$, $e_2$) are well determined from TTVs in spite of the non-transiting nature of the companion. The relationship between these parameters and observed TTV features can roughly be understood as follows, with the help of the analytic formula in Appendix \ref{app:formula}.

Most of the information comes from the spike-like feature around $\mathrm{BJD}\sim2454833+1050$, which is caused by the close encounter of the outer body and thus defines the time of its periastron passage $\tau_2$ (the effect is represented by $\mathcal{S}$ and $\mathcal{C}$ functions in Equations \ref{eq:sfunc} and \ref{eq:cfunc}; see also Figure \ref{fig:ttv_analytic}). In addition, its duration $\Delta t$ is sensitive to both $a_2$ and $e_2$: the former determines the overall orbital time scale and the latter changes the fraction of time spent around the periastron. 
The $\Delta t$ may be roughly estimated using Kepler's second law as follows:
\begin{align}
	\notag
	\Delta t&\sim\frac{P_2(1-e_2^2)^{3/2}}{2\pi}
	\int_{-\pi/2}^{+\pi/2} \frac{\mathrm{d}v_2}{(1+e_2\cos v_2)^2}\\
	\notag
	&\simeq \frac{P_2}{2}\left(1-\frac{4}{\pi}\,e_2\right)\\
	&\simeq \frac{P_1}{2}\left[ \frac{a_2(1-e_2)^{8/3\pi}}{a_1} \right]^{3/2}\sim\frac{P_1}{2}\left(\frac{q_2}{a_1}\right)^{3/2},
\end{align}
where its order of magnitude is not so sensitive to the rather arbitrary choice of the interval of integration ($-\pi/2$ to $\pi/2$).
Since $\Delta t\sim300\,\days$ and $P_1\sim20\,\days$, this estimate gives $q_2/a_1\sim10$. Finally, its amplitude $\Delta A$ is given by (Equations \ref{eq:quad}  and \ref{eq:quad_amp})
\begin{equation}
	\Delta A\simeq \frac{P_1}{2\pi}A_{\rm L1}\simeq P_1\frac{15}{16\pi}\left[\frac{q_2(1+e_2)}{a_1}\right]^{-3/2}\frac{m_\mathrm{c}}{m_\star},
\end{equation}
assuming $m_\mathrm{b}, m_\mathrm{c}\ll m_\star$. Combining $\Delta A\sim2\,\mathrm{min}$ with the above estimate $q_2/a_1\sim10$, we find $m_\mathrm{c}/m_\star\sim10^{-2}$. These numbers are in reasonable agreement with those from the full dynamical modeling.

In addition to the spike, a long-term modulation due to ``tidal delay" \citep[represented by $\mathcal{M}$ function in Equation \ref{eq:mfunc}]{2003A&A...398.1091B, 2005MNRAS.359..567A} is also visible (blue curves in Figure \ref{fig:ttv_analytic}): tidal force from the outer companion delays the inner binary period, depending on the distance between the two. This effect, combined with the short-term spike, allows for further constraints on $\tau_2$, $\omega_2$, $e_2$, and $P_2$. In principle, this additional constraint enables the separate determination of $a_2$ and $e_2$ rather than only $q_2$, although $a_2$ is not well constrained because it is not clear whether the whole cycle of the outer orbit is covered given the only one periastron passage observed.

The analytic expressions \ref{eq:quad}--\ref{eq:quad_noncopl} show that the short-term modulation represented by $\mathcal{S}$ or $\mathcal{C}$ function is produced only when $e_1\neq0$ or $i_{\rm 21}\neq0$. In fact, the observed TTVs alone are found to be fitted well either by the non-zero $e_1$ ($\delta_{\rm ecc}$ in Equation \ref{eq:quad}) or non-zero $i_{\rm 21}$ ($\delta_{\rm noncopl}$) effects. This causes the anti-correlated degeneracy between $e_1$ and $i_{\rm 21}$. Since $e_1$ also affects the spike amplitude, we have two classes of solutions: high-$e_1$--low-$i_{21}$--low-$m_\mathrm{c}$ solution and low-$e_1$--high-$i_{21}$--high-$m_\mathrm{c}$ solution. The joint TTV/TDV model slightly favors the latter.

The directions of the orbits, $\omega_1$, $\omega_2$,  $\cos i_2$, $\Omega_{21}$ are not well constrained from the TTVs alone. The complicated degeneracy between $\omega_2$ and $\Omega_{21}$ comes from the fact that TTVs are rather sensitive to such ``dynamical" angles referred to the invariant plane as $n_1$ and $n_2$ in figure 1 of \citet{2015MNRAS.448..946B}. 
Nevertheless, I use $\cos i_2$ and $\Omega_{21}$ because of the simplicity in implementing the isotropic prior. The relationship between the two sets of angles are summarized in Appendix \ref{app:conversion}.

Finally, the LTTE effect (Equation \ref{eq:ltte}) is about $0.1\,\mathrm{min}(a_2/\mathrm{au})$ for $m_\mathrm{c}/m_\star\sim10^{-2}$. This term therefore does not play a major role, as also seen in Figure \ref{fig:ttv_analytic}.

\subsection{Constraints from TTVs and TDVs}\label{ssec:koi12_joint}

In the joint analysis, I find a smaller $e_1$ than that derived from the TTVs alone, because the combination of $T$, $b$, and $\rho_\star$ favors a small $e_1\sin\omega_1$ \citep[see equations 16, 18, and 19 of][]{2011exop.book...55W}. This constraint is combined with those from TTVs to better determine $e_1$ and $\omega_1$. The distribution of $\omega_1$ is still bimodal because no constraint is available for $e_1\cos\omega_1$ from TDVs.

These additional constraints on $e_1$ and $\omega_1$ partly solve the degeneracies with other angles mentioned above. In particular, the value of $m_\mathrm{c}$ from the joint TTV/TDV analysis is larger than from the TTVs alone, because the low-$e_1$--high-$i_{21}$--high-$m_\mathrm{c}$ solution is favored.

Absence of significant TDVs only weakly constrains the mutual inclination. As mentioned, the solution with $i_{21}\sim90\degr$ is disfavored; if this were the case, a large tangential perturbation should have produced significant TDVs around the periastron passage, which are not present in the data. The solutions with $\Omega_{21}\sim0\degr$ and $\Omega_{21}\sim180\degr$ are basically indistinguishable and the log-likelihood is not sensitive to $\cos i_2$ either. In the TDV models (bottom left panel of Figure \ref{fig:bestfit_koi12}), both the positive and negative bumps are seen around the outer periastron because the inner inclination is perturbed in the opposite ways depending on the sign of $\cos i_2$. The current data do not favor either of the cases significantly. 

\section{Results: Kepler-693/KOI-824}\label{sec:koi824}

\begin{figure*}
	\centering
	\includegraphics[width=0.9\textwidth]{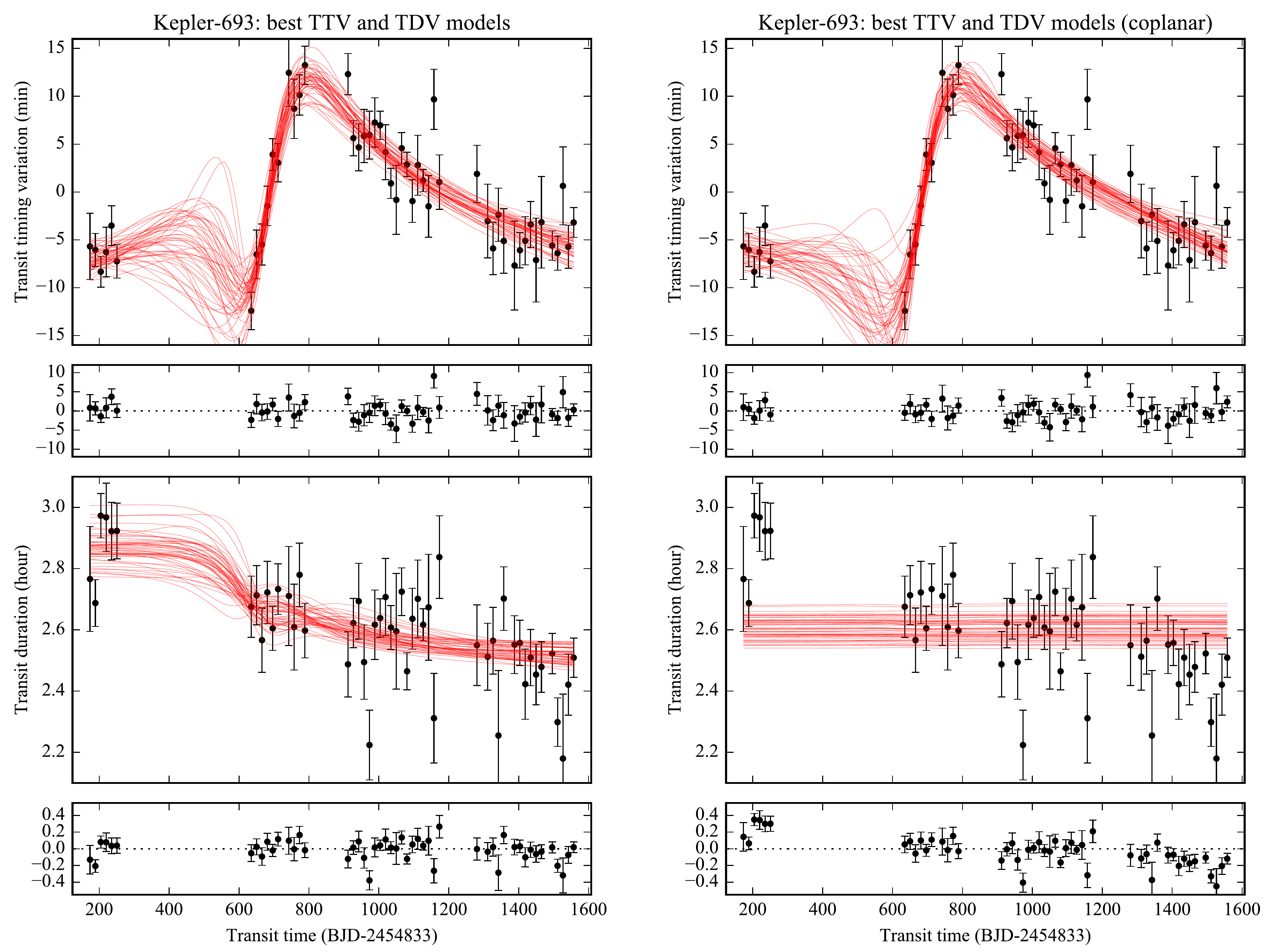}
	\caption{Dynamical models of the observed TTVs and TDVs of Kepler-693b.
	Red solid lines show $50$ models randomly sampled from
	the joint posterior distribution obtained by fitting the data
	(black circles with error bars). 
	The left column corresponds to the model which allows the mutual inclination to vary,
	while the right column shows the result when it is fixed to be zero.
	The coplanar model does not reproduce the observed transit durations.
	}
	\label{fig:bestfit_koi824}
\end{figure*}

The resulting posterior PDF from the {\tt MultiNest} analysis and the corresponding models are shown in Figures \ref{fig:bestfit_koi824} (red solid lines) and Figure \ref{fig:corner_koi824}. The summary statistics (median and $68\%$ credible region) of the marginal posteriors as well as the priors adopted for those parameters are given in Table \ref{tab:photod_koi824}. Again I found two almost identical solutions with prograde and retrograde orbits.

Similarly to Kepler-448c, the outer companion (we tentatively call it Kepler-693c) orbits close to the inner WJ in a highly eccentric orbit, except that it has a larger mass $1.5^{+0.6}_{-0.4}\times10^2\,M_{\rm Jup}$ and can be a low-mass star. The outer pericenter distance of $1.5\pm0.2\,\mathrm{au}$ is also similar to Kepler-448c and satisfies the stability condition \citep{2001MNRAS.321..398M}. Again non-zero eccentricity is favored for the inner orbit ($e_1=0.2^{+0.2}_{-0.1}$), and its mass is not constrained at all from the data.

A notable difference from the previous case is the clear TDV signal, which tightly constrains the mutual inclination to be $|i_{21}-90^\circ|\simeq40\degr$ (Figure \ref{fig:imut}, bottom). I checked that the data can never be explained by the aligned configuration (right column of Figure \ref{fig:bestfit_koi824}): I find the Bayesian evidence for the mutually-inclined model ($\ln\mathcal{Z}=-89.18\pm0.08$) is larger than that of the coplanar one ($\ln\mathcal{Z}_{\rm copl}=-103.1\pm0.07$) by a factor of $10^6$. 
The observed mutual inclination is the largest among those dynamically measured for planetary systems, and likely above the critical angle for the Kozai oscillation (posterior probability that $39\fdg2<i_{21}<140\fdg8$ is $80\%$).\footnote{\citet{2014A&A...571A..37S} inferred the orbital inclination of the binary companion of the transiting WJ host KOI-1257/Kepler-420 to be $i_2=18\fdg2^{+18\fdg0}_{-5\fdg4}$ ($68.3\%$ interval), based on the trend in the SOPHIE RVs and bisector and full-width half maximum of the stellar lines, combined with the {\it Kepler} light curve and stellar spectral energy distribution. If confirmed, the value translates into a large $i_{21}\gtrsim90\degr-i_2$ ($\Omega_{21}$ is not constrained at all) and a similar eccentricity evolution as discussed in the present paper is expected, although they caution that the outer binary inclination is poorly constrained, with the $99\%$ limit on $i_2$ being $[8\fdg2, 85\fdg2]$.} In Section \ref{sec:secular}, I will show that the perturbation from the inclined companion does cause a large eccentricity oscillation of the inner WJ. 

Kepler-693b's mass is not measured, and the planet was statistically validated by \citet{2016ApJ...822...86M}. A possible concern is that the absence of secondary eclipse, which the statistical validation partly relies on \citep[in addition to other factors including the transit shape and non-detection of the companion via the AO imaging by][]{2016AJ....152...18B}, may lose its meaning if $b_{\rm occ}$ is larger than unity due to the non-zero inner eccentricity. However, our dynamical modeling finds $b_{\rm occ}=0.64^{+0.08}_{-0.09}$, which excludes the possibility.
In addition, the derived mean stellar density is also compatible with the KIC value. Therefore, the low false positive probability (less than $1/3000$) derived by \citet{2016ApJ...822...86M} is still valid in the light of our new constraints. 

\begin{deluxetable*}{lcccc}
	\tablewidth{0pt}
	\tablecolumns{2}
	\tablecaption{Parameters of the Kepler-693 System from the Dynamical TTV and TDV Analysis\label{tab:photod_koi824}}
	\tablehead{
		\colhead{Parameter} & \colhead{Solution 1} & \colhead{Solution 2} 
		& \colhead{Combined} & \colhead{Prior}
	}
	\startdata
	{\bf Fitted Parameters}\\
	({\it Inner Orbit})\\
	1. Time of inferior conjunction
		& $173.611^{(+1)}_{(-1)}$ & $173.610^{(+2)}_{(-1)}$ & $173.610^{(+1)}_{(-1)}$
		& $\uni(173.58, 173.64)$\\
		\quad $t_{\rm ic, 1}$ ($\kbjd$)  &&&\\
	2. Orbital period $P_1$ (day)
		& $15.37541^{(+10)}_{(-7)}$ & $15.37534^{(+10)}_{(-8)}$
		& $15.37537^{(+10)}_{(-9)}$
		& $\luni(15.375, 15.376)$\\
	3. Orbital eccentricity $e_1$
		& $0.14^{+0.08}_{-0.04}$ & $0.3^{+0.2}_{-0.1}$ & $0.2^{+0.2}_{-0.1}$
		& $\uni(0, 0.7)$\\
	4. Argument of periastron $\omega_1$ (deg)
		& $41^{+55}_{-28}$ & $174^{+13}_{-39}$ & \nodata
		& $\uni(-180, 180)$\\
	5. Cosine of orbital inclination $\cos i_1$\tablenotemark{a}
		& $0.021(2)$ & $0.022(3)$ & $0.022(3)$
		& $\uni(0, 0.04)$\tablenotemark{a}\\
	({\it Outer Orbit})\\
	6. Time of the periastron passage 
		& $640^{+20}_{-17}$ & $640^{+22}_{-18}$ & $640^{+22}_{-17}$
		& $\uni(500, 900)$\\
		\quad $\tau_2$ ($\kbjd$) &&&\\
	7. Periastron distance over 
		& $13^{+2}_{-1}$ & $13^{+1}_{-2}$ & $13^{+2}_{-2}$
		& $\luni(6, 25)$\\
		\quad inner semi-major axis $q_2/a_1$ &&&\\
	8. Orbital eccentricity $e_2$
		& $0.48^{+0.12}_{-0.06}$ & $0.46^{+0.10}_{-0.05}$ & $0.47^{+0.11}_{-0.06}$
		& $\uni(0, 0.95)$\\
	9. Argument of periastron $\omega_2$ (deg)
		& $41^{+20}_{-20}$ & $25^{+25}_{-21}$ & $30^{+24}_{-23}$
		& $\uni(-180, 180)$\\
	10. Cosine of orbital inclination $\cos i_2$\tablenotemark{a}
		& $-0.2^{+0.3}_{-0.2}$ & $-0.3^{+0.2}_{-0.1}$ & $-0.3^{+0.2}_{-0.2}$
		& $\uni(-1, 1)$\\
	11. Relative longitude of 
		& $51^{+8}_{-10}$ & $-138^{+12}_{-11}$ & \nodata
		& $\uni(-180, 180)$\\
		\quad ascending node $\Omega_{21}$ (deg)\tablenotemark{a,b} &&&\\
	({\it Physical Properties})\\
	12. Mass of Kepler-693 $m_\star$ ($M_\odot$)
		& $0.80^{+0.03}_{-0.03}$	& $0.80^{+0.04}_{-0.03}$ & $0.80^{+0.04}_{-0.03}$
		& $\gaus(0.79, 0.03, 0.15)$\\
	13. Mean density of 
		& $2.2^{+0.3}_{-0.2}$	& $2.2^{+0.3}_{-0.3}$ & $2.2^{+0.3}_{-0.3}$
		& $\gaus(1.93, 0.18, 0.53)$\\
		\quad Kepler-693 $\rho_\star$ ($\gcc$) &&&\\
	14. Mass of Kepler-693b $m_\mathrm{b}$ ($M_{\rm Jup}$)\tablenotemark{c}
		& $0.8^{+2.7}_{-0.6}$ & $1.0^{+3.2}_{-0.7}$ & $0.9^{+3.0}_{-0.7}$
		& $\luni(0.1, 10)$\\
	15. Mass of Kepler-693c $m_\mathrm{c}$ ($M_{\rm Jup}$)
		& $167^{+59}_{-43}$ & $136^{+50}_{-34}$ & $145^{+58}_{-37}$
		& $\luni(0.001M_\odot, 0.3M_\odot)$\\
	({\it Jitters})\\
	16. Transit time jitter $\sigma_{\rm TTV}$ ($10^{-4}\,\mathrm{day}$)
		& $2^{+4}_{-1}$		& $1^{+4}_{-1}$ & $1^{+4}_{-1}$
		& $\luni(5\times10^{-2}, 50)$\\
	17. Transit duration jitter $\sigma_{\rm dur}$ ($10^{-4}\,\mathrm{day}$)
		& $9^{+14}_{-8}$ & $9^{+13}_{-8}$ & $9^{+14}_{-8}$
		& $\luni(0.1, 10^{2})$\\
	\\
	{\bf Derived Parameters}\\
	Outer orbital period $P_2$ (day)
		& $(1.8^{+1.0}_{-0.4})\times10^3$ & $(1.8^{+0.6}_{-0.3})\times10^3$
		& $(1.8^{+0.8}_{-0.3})\times10^3$
		& \nodata\\
	Inner semi-major axis $a_1$ (au)
		&  $0.112^{(+2)}_{(-1)}$ & $0.112^{(+2)}_{(-2)}$ 
		& $0.112^{(+2)}_{(-1)}$ & \nodata\\
	Outer semi-major axis $a_2$ (au)
		&  $2.9^{+1.1}_{-0.5}$ & $2.8^{+0.7}_{-0.4}$ 
		& $2.8^{+0.8}_{-0.4}$ & \nodata\\
	Periastron distance of 
		&  $1.5\pm0.2$ &  $1.5\pm0.2$ &  $1.5\pm0.2$ & \nodata\\
		\quad the outer orbit $a_2(1-e_2)$ (au) &&& \\
	Mutual orbital inclination $i_{21}$ (deg)
		& $53^{+7}_{-9}$	& $134^{+11}_{-10}$	
		& \nodata & \nodata\\
	Physical radius of Kepler-693 ($R_\odot$)
		& $0.80\pm0.03$	& $0.80\pm0.04$ & $0.80\pm0.04$ & \nodata\\
	Physical radius of Kepler-693b ($R_\mathrm{Jup}$)\tablenotemark{d}
		& $0.91\pm0.05$	& $0.91\pm0.05$ 	& $0.91\pm0.05$ & \nodata\\
	Transit impact parameter of Kepler-693b
		& $0.58\pm0.04$ & $0.59\pm0.05$ & $0.59\pm0.05$ & \nodata\\
	Occultation impact parameter of Kepler-693b
		& $0.68^{+0.06}_{-0.06}$ & $0.62^{+0.09}_{-0.08}$ & $0.64^{+0.08}_{-0.09}$ & \nodata\\
	\\
	Log evidence $\ln \mathcal{Z}$ from {\tt Multinest} & $-90.30\pm0.08$ & $-89.57\pm0.08$ & $-89.18\pm0.08$\\
	\enddata
	\tablecomments{The elements of the inner and outer orbits listed here
	are Jacobian osculating elements defined at the epoch 
	$\mathrm{BJD}=2454833+170$.
	The quoted values in the `Solution' columns are the median
	and $68\%$ credible interval of the marginal posteriors.
	Parentheses after values denote uncertainties in the last digit.
	The `combined' column shows the values from the marginal posterior 
	combining the two solutions; no value is shown if the combined 
	marginal posterior is multimodal. In the prior column,
	$\uni(a, b)$ and $\luni(a, b)$ denote the (log-)uniform priors between $a$ and $b$,
	$f(x)=1/(b-a)$ and $f(x)=x^{-1}/(\ln b - \ln a)$, respectively;
	$\gaus(a, b, c)$ means the asymmetric Gaussian prior 
	with the central value $a$ and lower and upper widths $b$ and $c$.}
	\tablenotetext{a}{\red{There also exists a solution with negative $\cos i_1$. The solution is statistically indistinguishable from the one reported here, except that the signs of $\cos i_2$ and $\Omega_{21}$ are flipped (and thus $i_{21}$ remains the same; see Eqn. \ref{eq:i21}). In principle, the TTVs for the solutions with $\cos i_1>0$ and $\cos i_1<0$ are not completely identical \citep[see, e.g., A15 of][]{2015MNRAS.448..946B}. However, the difference is in practice negligibly small for a transiting system because the effect is proportional to $\cot i_1$. I confirmed that this is indeed the case by performing the same numerical fit with the prior on $\cos i_1$ replaced by $\uni(-0.04, 0)$.}}
	\tablenotetext{b}{Referenced to the ascending node of the inner orbit,
	whose direction is arbitrary.}
	\tablenotetext{c}{This parameter is not constrained by the data; posterior is identical to the log-uniform prior.}
	\tablenotetext{d}{Derived from the posterior of $R_\star$ and that of  $R_{\rm p}/R_\star$ from the mean transit light curve (Table \ref{tab:tpars_koi824}).}
\end{deluxetable*}

\subsection{Adopted Parameters}

I adopt $m_\star=0.793^{+0.054}_{-0.029}\,M_\odot$ and $\rho_\star=1.931^{+0.526}_{-0.1773}\,\gcc$ (corresponding to $r_\star=0.833^{+0.033}_{-0.062}\,R_\odot$) from KIC 
as the priors. I adopt $b_{\rm mean}=0.6$ and $\sigma_b=0.1$ that roughly incorporates the $68\%$ credible region of the posterior from the mean transit light curve (Table \ref{tab:tpars_koi824}).

\subsection{Constraints from TTVs}

The situation is basically similar to the Kepler-448 case, except that the TTV amplitude is larger by almost an order of magnitude and so is the companion's mass. In addition, the whole shape of the short-term feature is not entirely observed due to the data gap, as shown in the large scatter of the models (Figure \ref{fig:bestfit_koi824}). This explains a weaker constraint on $q_2/a_1$, which is mainly determined by the duration of the feature (Section \ref{ssec:koi12_ttv}).

\subsection{Constraints from TTVs and TDVs}

It is also the case in this system that the duration data point to a lower inner eccentricity than favored by the TTVs alone and increase the estimated mass. A striking difference is the clear trend in the duration that points to non-zero $i_{21}$. The trend comes from the nodal precession of the inner orbit \citep{2002ApJ...564.1019M} and has also been observed in the Kepler-108 system recently \citep{2017AJ....153...45M}. While a similar duration drift can also be caused by the apsidal precession of the eccentric inner orbit, this is unlikely to be the case given the failure of the coplanar model. In fact, both precession frequencies are basically comparable, but the effect of the apsidal precession on the duration is smaller than that of the nodal precession by a factor of $R_\star/a_1$ \citep{2002ApJ...564.1019M}. This difference can be significant for WJs with $a/R_\star>10$. 

\subsubsection{Mutual Inclination from the Likelihood Profile}

\red{
The constraint on the mutual inclination in Figure \ref{fig:imut} and Table \ref{tab:photod_koi824} is based on the one-dimensional Bayesian posterior marginalized over the other parameters. This depends not only on the likelihood and prior function but also on volume of the parameter space with high posterior probabilities. Thus, the resulting credible interval is generally not the same as the likelihood-based confidence interval even for a uniform prior, and a low value in the marginal posterior probability does not necessarily mean a poor fit to the data \citep{2011JHEP...06..042F}.}

To see what value of the mutual inclination is ``excluded" by the data alone in the frequentist sense, I also estimate the confidence interval for $i_{21}$ based on the likelihood profile. Specifically, I derive the maximum likelihood $\hat{\mathcal{L}}(i_{21})$ for each fixed value of $i_{21}$ optimizing the other parameters, and examine its form as a function of $i_{21}$. This has been done by searching for minimum $\chi^2$ solutions for a grid of $\cos i_2\in[-1,1]$ and $\Omega_{21}\in[-90\degr, 90\degr]$ (prograde solutions). Here the constraints on $m_\star$ and $\rho_\star$ are also incorporated in $\chi^2$, and $\sigma_{\rm TTV}$ and $\sigma_{\rm dur}$ were fixed to be zero.

\begin{figure}
	\centering
	\includegraphics[width=\columnwidth]{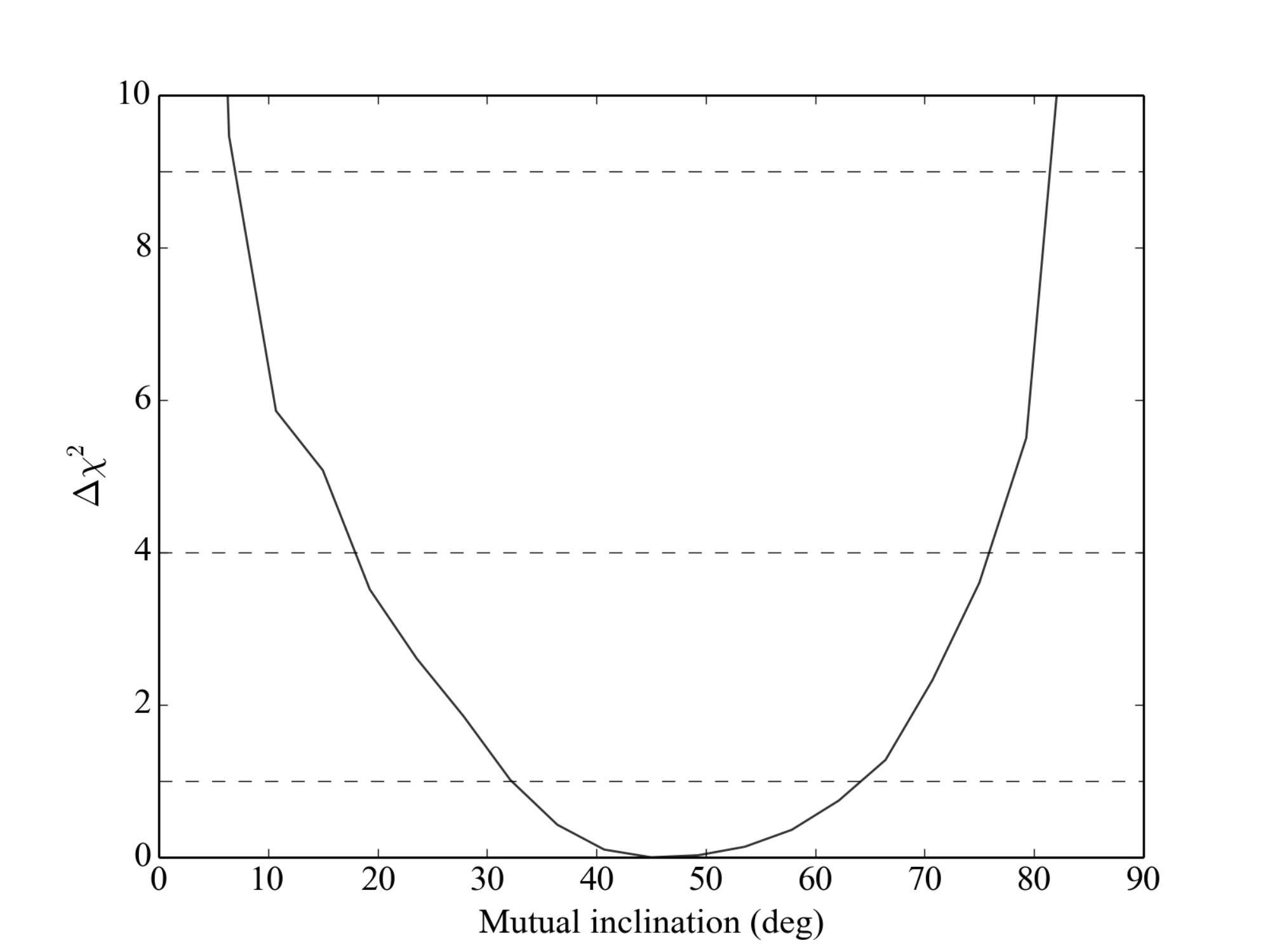}
	\caption{Chi-squared difference from the global best-fit value as a function of the mutual inclination between the inner and outer orbits. The horizontal dashed lines show $\Delta\chi^2=1$, $4$, and $9$, which correspond to $1\sigma$, $2\sigma$, and $3\sigma$ confidence intervals, respectively.
	}
	\label{fig:chi2_imut}
\end{figure}

Figure \ref{fig:chi2_imut} shows the resulting profile of $-2\ln(\hat{\mathcal{L}}(i_{21})/\hat{\hat{\mathcal{L}}})$, where $\hat{\hat{\mathcal{L}}}$ denotes the maximum likelihood found by optimizing all the model parameters including $i_{21}$. This is equivalent to the chi-squared difference from its minimum value, $\Delta \chi^2$, in our current setting. Here the $\Delta\chi^2$ value is scaled so that the minimum $\chi^2$ solution has $\chi^2_\nu=1$. The resulting $1\sigma$ ``confidence" interval is $i_{21}=47\degr^{+17\degr}_{-15\degr}$, 
and the $2\sigma$ lower bound is found to be $i_{21}=18\degr$. Thus I conclude that a high mutual inclination is indeed favored by the data and the conclusion is insensitive to the prior information on $i_{21}$.

\section{Long-term Orbital Evolution}\label{sec:secular}

A high mutual inclination confirmed for the Kepler-693 system may result in a large eccentricity oscillation of the inner orbit as long as the perturbation is strong enough to overcome other short-range forces. If this is indeed the case, the system may serve as a direct piece of evidence that some WJs are undergoing eccentricity oscillations. Even in the Kepler-448 system where highly-inclined solutions are not necessarily favored, significant eccentricities of both the inner and outer orbits may still lead to the eccentricity excitation due to the octupole-level interaction. 

Given the full set of parameters constrained from the dynamical analysis, 
\red{we can assess the future of the systems rather reliably by extrapolating the dynamical solutions.}
In this section, we explore the effect of the secular perturbation due to the outer companions on the inner planet's orbit and its tidal evolution.

\subsection{Oscillation of the Inner Eccentricity}\label{ssec:secular_ecc}

I compute secular evolution of both the inner and outer orbits along with the spins of the star and inner planet. I use the code developed and utilized in \citet{2016ApJ...820...55X} and \citet{2017ApJ...835..204X}, which takes into account (i) gravitational interaction up to the octupole order and (ii) precessions due to general relativity as well as tidal and rotational deformation of the bodies. Here I neglect magnetic braking of the star and tidal dissipation inside the bodies, and assume zero stellar and planetary obliquities for simplicity. The rotation periods are set to be $1.25\,\mathrm{days}$ for Kepler-448, $10\,\mathrm{days}$ for Kepler-693, and $1\,\mathrm{day}$ for the inner planets, although the spin evolution does not affect the result significantly. I adopt standard values for the dimensionless moments of inertia (0.059 and 0.25 for the star and inner planet) and the tidal Love numbers (0.028 and 0.5). The orbits are integrated for $10\,\mathrm{Myr}$ (sufficiently longer than the oscillation timescale; see below), starting from $1000$ random sets of parameters sampled from the posterior distribution from the dynamical analyses (Sections \ref{sec:koi12} and \ref{sec:koi824}). 

\begin{figure*}
	\centering
	\fig{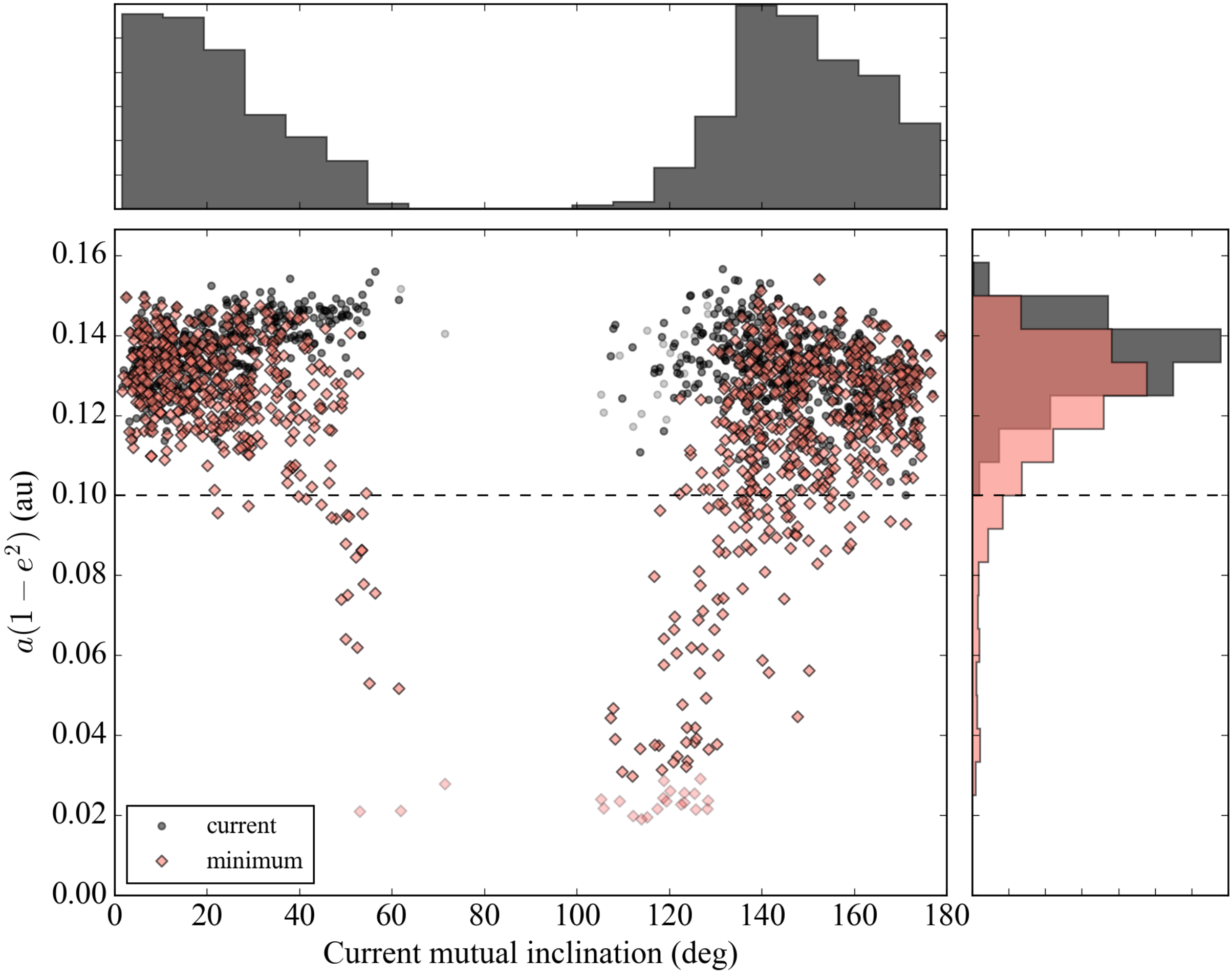}{0.49\textwidth}{(a) Kepler-448}
	\fig{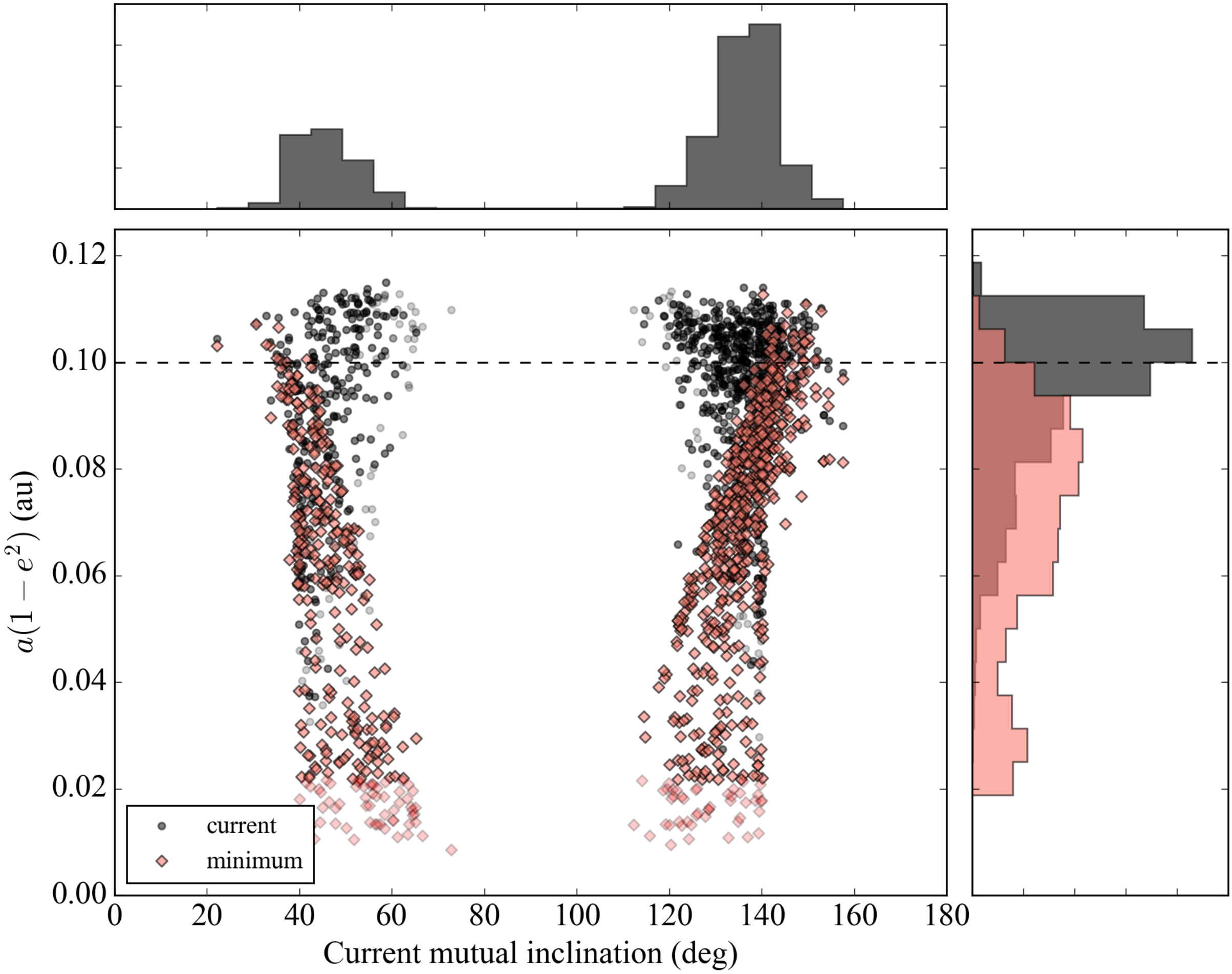}{0.49\textwidth}{(b) Kepler-693}
	\caption{Effect of secular eccentricity oscillation on the semi-latus rectum of the inner orbit, $a_1(1-e_1^2)$. This distance corresponds to the final semi-major axis of the HJ if the orbit is circularized at a fixed angular momentum. The black circles and histograms are randomly sampled from the posterior of the dynamical TTV/TDV modeling and represent the current values. The red diamonds and histograms are the minimum values over many oscillation timescales (Section \ref{ssec:secular_ecc}). Solutions with minimum periastron distance $a_1(1-e_1)$ less than the Roche limit \citep[here chosen to be $2.7\,R_\star$ based on][simply assuming the planetary mass $10^{-3}\,M_\star$ and radius $0.1\,R_\star$]{2011ApJ...732...74G} are plotted with transparent colors and not included in the histogram. The horizontal dashed line indicates $a_1(1-e_1^2)=0.1\,\mathrm{au}$, which is the conventional threshold for possible tidal circularization accepted by \citet{2012ApJ...750..106S}, \citet{2014ApJ...781L...5D}, and \citet{2015ApJ...798...66D}.
	}
	\label{fig:secular_ecc}
\end{figure*}

Figure \ref{fig:secular_ecc} shows the initial (black circles) and minimum (red diamonds) values of $a_1(1-e_1^2)$ over the course of evolution, the latter of which correspond to the maxima of $e_1$. In both systems, significant eccentricity oscillations occur at least for some of the solutions. 
If we adopt $a(1-e^2)<0.1\,{\rm au}$ as a conventional threshold for the migrating WJs \citep{2012ApJ...750..106S, 2014ApJ...781L...5D, 2015ApJ...798...66D}, the periastrons become close enough to drive the tidal migration for $12\%$ of the solutions for Kepler-448b and $96\%$ for Kepler-693b, excluding the tidally-disrupted cases shown with transparent colors. If we choose $a(1-e^2)<0.05\,{\rm au}$ instead as a threshold \citep[e.g.,][]{2016MNRAS.456.3671A}, the fractions become $5\%$ and $33\%$ for Kepler-448b and Kepler-693b, respectively. Large eccentricity oscillations (and hence small minimum periastrons) are observed mainly for $40\degr\lesssim i_{21} \lesssim140\degr$; this explains why a larger fraction of solutions have sufficiently small $a(1-e^2)$ for the Kepler-693 system. 

\begin{figure*}
	\centering
	\fig{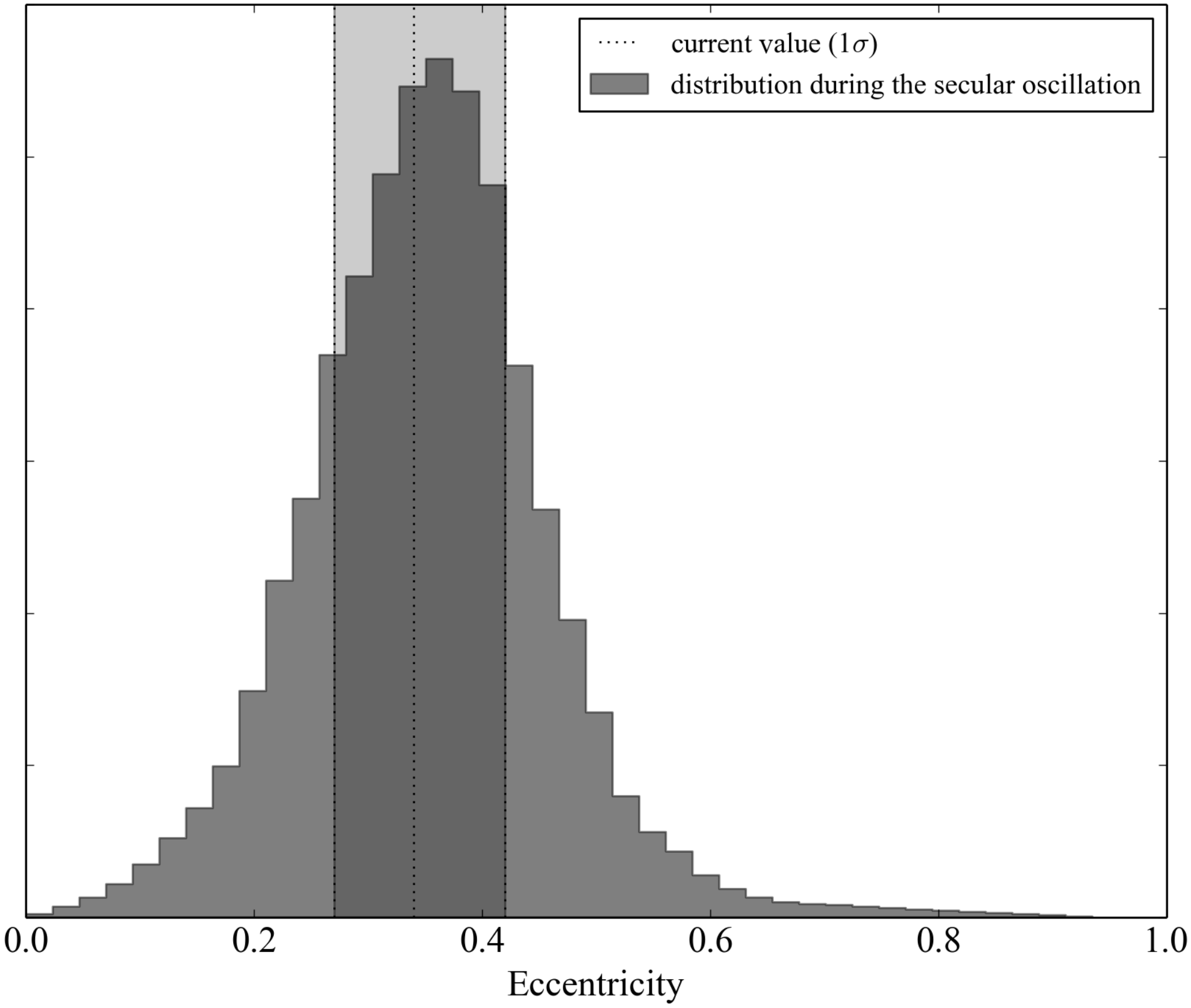}{0.43\textwidth}{(a) Kepler-448b}
	\fig{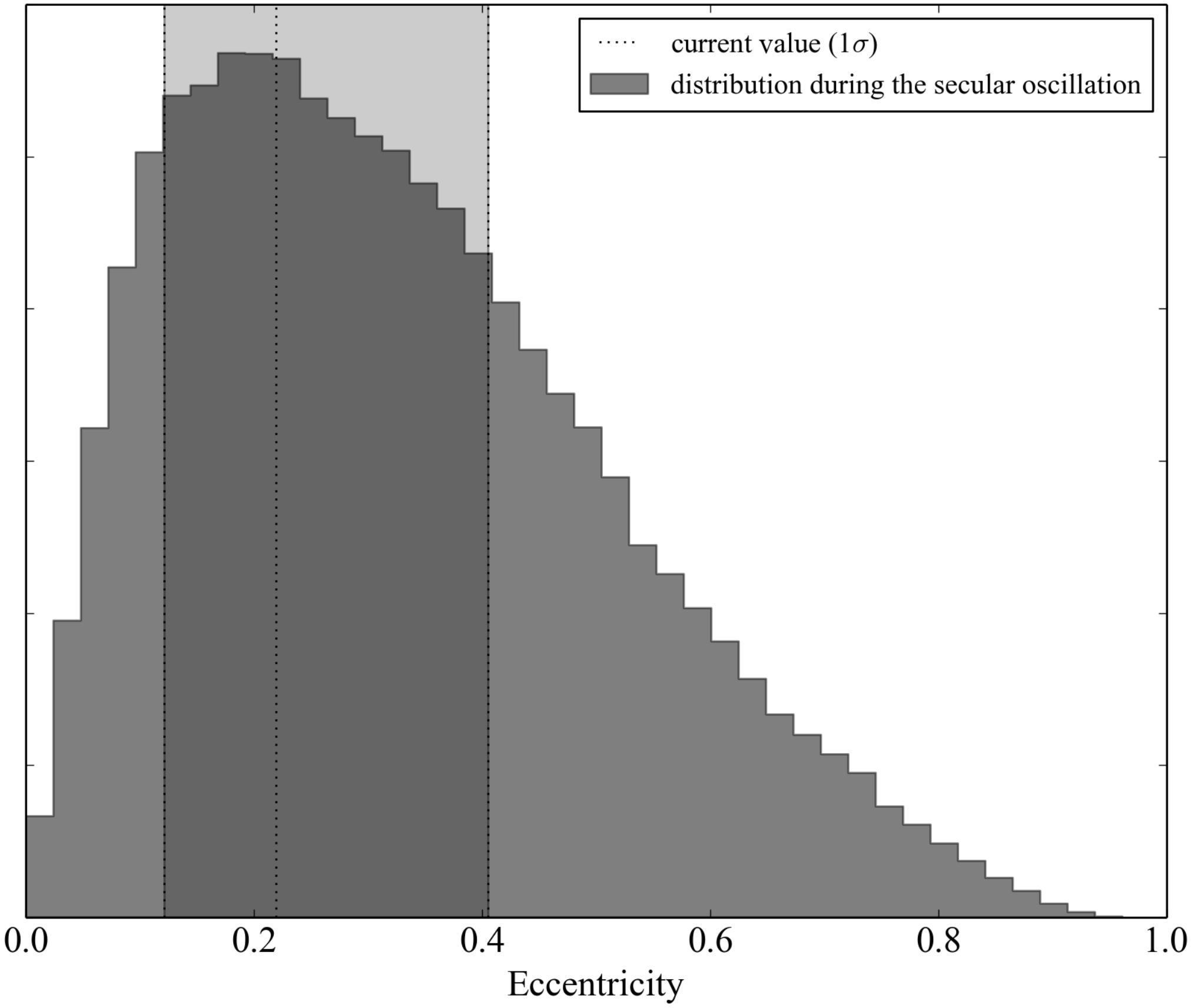}{0.43\textwidth}{(b) Kepler-693b}
	\caption{The distributions of the inner eccentricities ($e_1$) during the secular oscillation averaged over the dynamical solutions. The vertical dotted lines (shaded region) show the median and $68\%$ credible interval of the current $e_1$ measured from the TTVs and TDVs.
	}
	\label{fig:ecc_dist}
\end{figure*}

The current eccentricities of the two WJs ($1\sigma$ region shown with vertical dotted lines in Figure \ref{fig:ecc_dist}) also turn out to be the most likely values to be observed over the course of oscillation if we are random observers in time; they are around the peaks of the inner eccentricity distribution during the $10\,\mathrm{Myr}$ evolution averaged over the dynamical solutions (gray histograms in Figure \ref{fig:ecc_dist}).

For highly inclined solutions, both libration and circulation of the argument of periastron \citep{1962AJ.....67..591K}, modified by the octupole effects, are observed.\footnote{On the other hand, the libration of the difference in the apsidal longitudes $\Delta \varpi_{\rm inv}$,
as discussed in \citet{2014Sci...346..212D}, was not observed in the current simulations.} 
In either of the two systems, the Kozai-Lidov timescale $\tau_{\rm KL}$ \citep{1998MNRAS.300..292K} and the octupole one $\tau_{\rm Oct}$ are both found to be much shorter than that of general relativistic apsidal precession $\tau_{\rm GR}$  for any solution found from TTVs and TDVs: for reference, I find typical $\tau_{\rm KL}\sim10^{-2}\,\mathrm{Myr}$, $\tau_{\rm Oct}\sim10^{-1}\,\mathrm{Myr}$, and $\tau_{\rm GR}\sim10^{0.5}\,\mathrm{Myr}$ for Kepler-448, and $\tau_{\rm KL}\sim10^{-3}\,\mathrm{Myr}$, $\tau_{\rm Oct}\sim10^{-1.5}\,\mathrm{Myr}$, and $\tau_{\rm GR}\sim10^{0.5}\,\mathrm{Myr}$ for Kepler-693. Thus the perturbation from the ``close friends" is indeed strong enough to overcome general relativity \citep{2014ApJ...781L...5D}. Note that this is the case for the octupole effect as well, which explains why some low-$i_{21}$ solutions in the Kepler-448 system lead to a large eccentricity oscillation. 

\subsection{Migration Timescales}\label{ssec:secular_tau}

Are the two WJs currently migrating into HJs due to the eccentricity oscillation? If this is the case, the current migration timescale $\tau_{\rm mig}$ needs to be comparable to the system age $\tau_{\rm age}$. If $\tau_{\rm mig}\gg\tau_{\rm age}$ they are unlikely to be migrating, while if $\tau_{\rm mig}\ll\tau_{\rm age}$ the WJs should have evolved into HJs rapidly and we are unlikely to observe the system in the current state. Here we perform an order-of-magnitude comparison between the two timescales, given their large observational and theoretical uncertainties.

\begin{figure*}
	\centering
	\fig{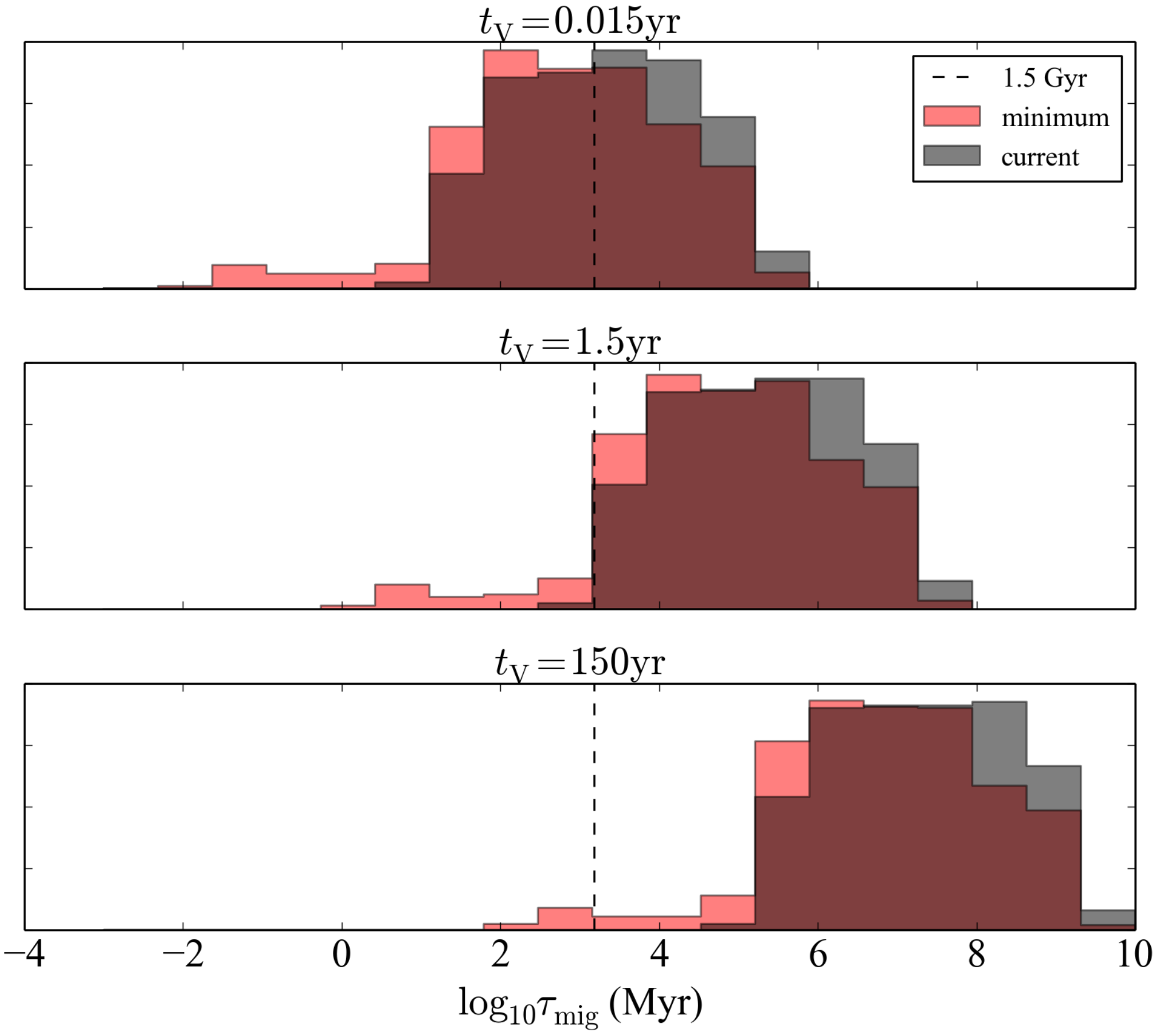}{0.49\textwidth}{(a) Kepler-448}
	\fig{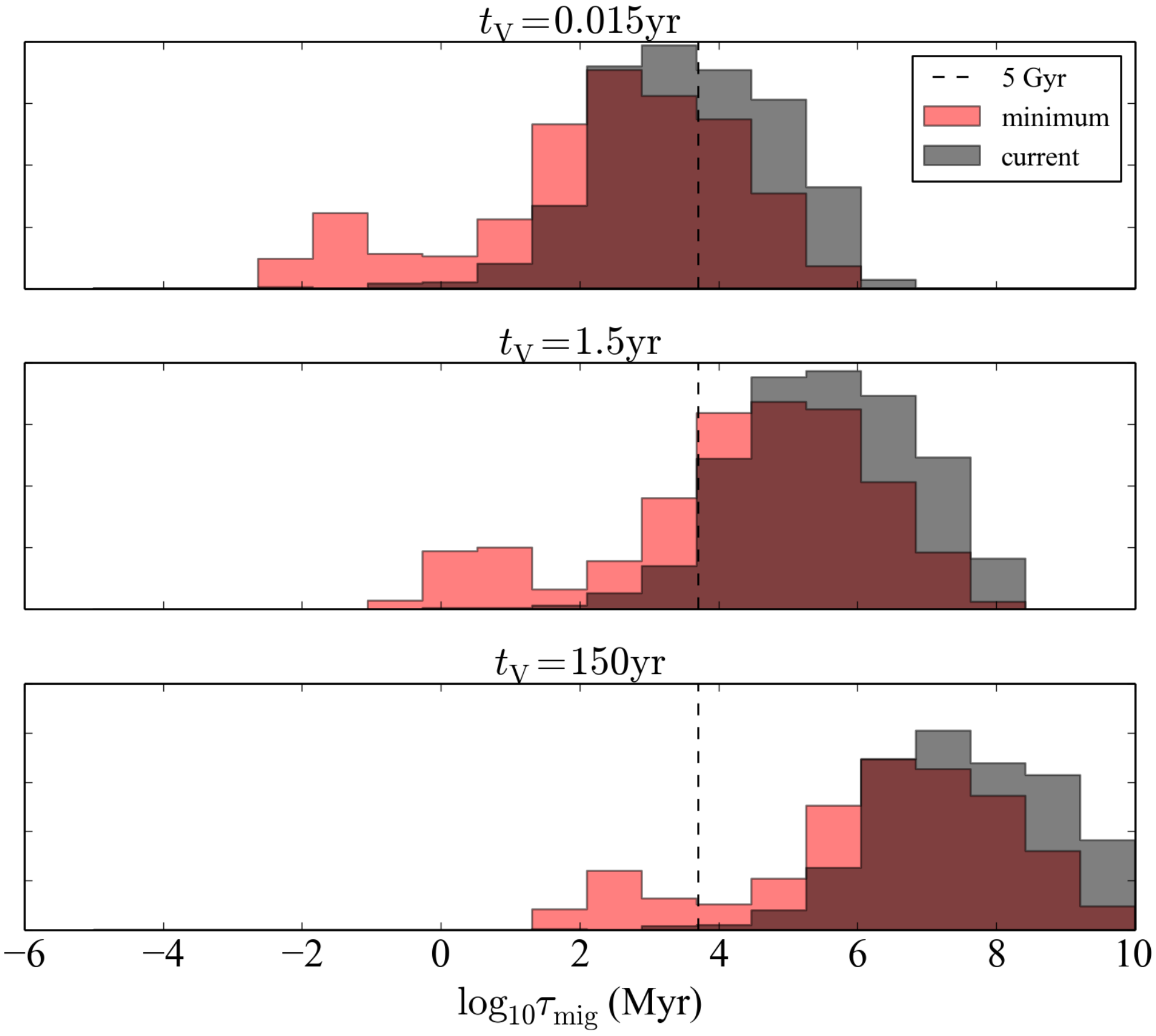}{0.49\textwidth}{(b) Kepler-693}
	\caption{Migration timescales of the inner WJs for three different viscous timescales $t_{\rm V}$, computed with equation 2 of \citet{2016ApJ...829..132P}. The red and black histograms show the values at the minimum periastron over the course of eccentricity oscillation (Section \ref{ssec:secular_ecc} and Figure \ref{fig:secular_ecc}) and at the initial (current) periastron, respectively. The vertical dashed lines indicate the current age of the system estimated from spectroscopy (for Kepler-448) and speculated from the host-star property (for Kepler-693).
	}
	\label{fig:tau}
\end{figure*}

Figure \ref{fig:tau} shows the distribution of tidal migration timescale given by equation 2 of \citet{2016ApJ...829..132P} at the minimum periastron distance computed in Section \ref{ssec:secular_ecc}. Strictly speaking, the relevant timescale is expected to be longer because oscillation of $e_1$ slows down the migration \citep{2015ApJ...799...27P}, but in our case the maximum eccentricity is not so close to unity (since $a_1$ is already small) that the modification is unlikely large \citep[\textit{cf.} figure 2 of][]{2015ApJ...799...27P}; thus we simply neglect the correction. The timescales are computed for \red{three different values of the viscous time of the planet, which gives a characteristic timescale for dissipation: $t_{\rm V}=0.015\,\mathrm{yr}$, $t_{\rm V}=1.5\,\mathrm{yr}$, and $t_{\rm V}=150\,\mathrm{yr}$, while the dissipation inside the star is neglected (see \citet{2012arXiv1209.5724S} for comparison with different parameterizations). The three values} roughly correspond to (i) very efficient tidal dissipation required for some high-eccentricity migration scenarios to explain the observed HJs \citep{2015ApJ...799...27P, 2017MNRAS.464..688H}, (ii) dissipation required to circularize the orbits of HJs with $P\leq5\,\mathrm{days}$ within $10\,\mathrm{Gyr}$ \citep{2012arXiv1209.5724S}, and (iii) values calibrated based on a sample of eccentric planetary systems \citep{2010ApJ...723..285H, 2014ApJ...787...27Q}, while the limit from the Jupiter--Io system ($t_{\rm V}\gtrsim15\,\mathrm{yr}$) lies in between the latter two. I also indicate (rough) estimates for the ages of the two systems with vertical dashed lines: $1.5\,\mathrm{Gyr}$ for Kepler-448 based on spectroscopy \citep{2015A&A...579A..55B} and $5\,\mathrm{Gyr}$ for Kepler-693 as a tentative value given that the host star has dimensions of a K dwarf.

The comparison between the histogram and the dashed line shows that the migrating solutions with $\tau_{\rm mig}\sim\tau_{\rm age}$ exist for a wide range of $t_{\rm V}$ for the Kepler-693 system. In particular, the eccentricity oscillation plays a crucial role for $t_{\rm V}\gtrsim1.5\,\mathrm{yr}$ so that such solutions realize. The migrating solutions also exist for the Kepler-448 system, though they seem plausible only for a small fraction of significantly misaligned solutions that lead to the large eccentricity oscillation, or require efficient tidal dissipation with $t_{\rm V}\lesssim1.5\,\mathrm{yr}$.




\subsubsection{Possible Fates of the Inner Planets}

The timescale arguments above indicate the inner WJs may 
evolve into HJs within the lifetime of the system. If this is the case, HJ systems with close substellar companions as found by a long-term RV monitoring \citep{2017MNRAS.tmp..167T} may have been WJs like ours in the past.

As a proof of concept, I compute the evolution of the two systems for $1\,\mathrm{Gyr}$, including the tidal dissipation with the planetary viscous timescale $t_{\rm V}=1.5\,\mathrm{yr}$ for an illustration. I fix $t_{\rm V}=50\,\mathrm{yr}$ for the star. I stop the calculation when the inner orbit is circularized (both $a_1<0.1\,\mathrm{au}$ and $e_1<0.01$ are achieved) before $1\,\mathrm{Gyr}$. If we change $t_{\rm V}$, we expect things just happen on a different timescale corresponding to the change. 

\begin{figure*}
	\centering
	\fig{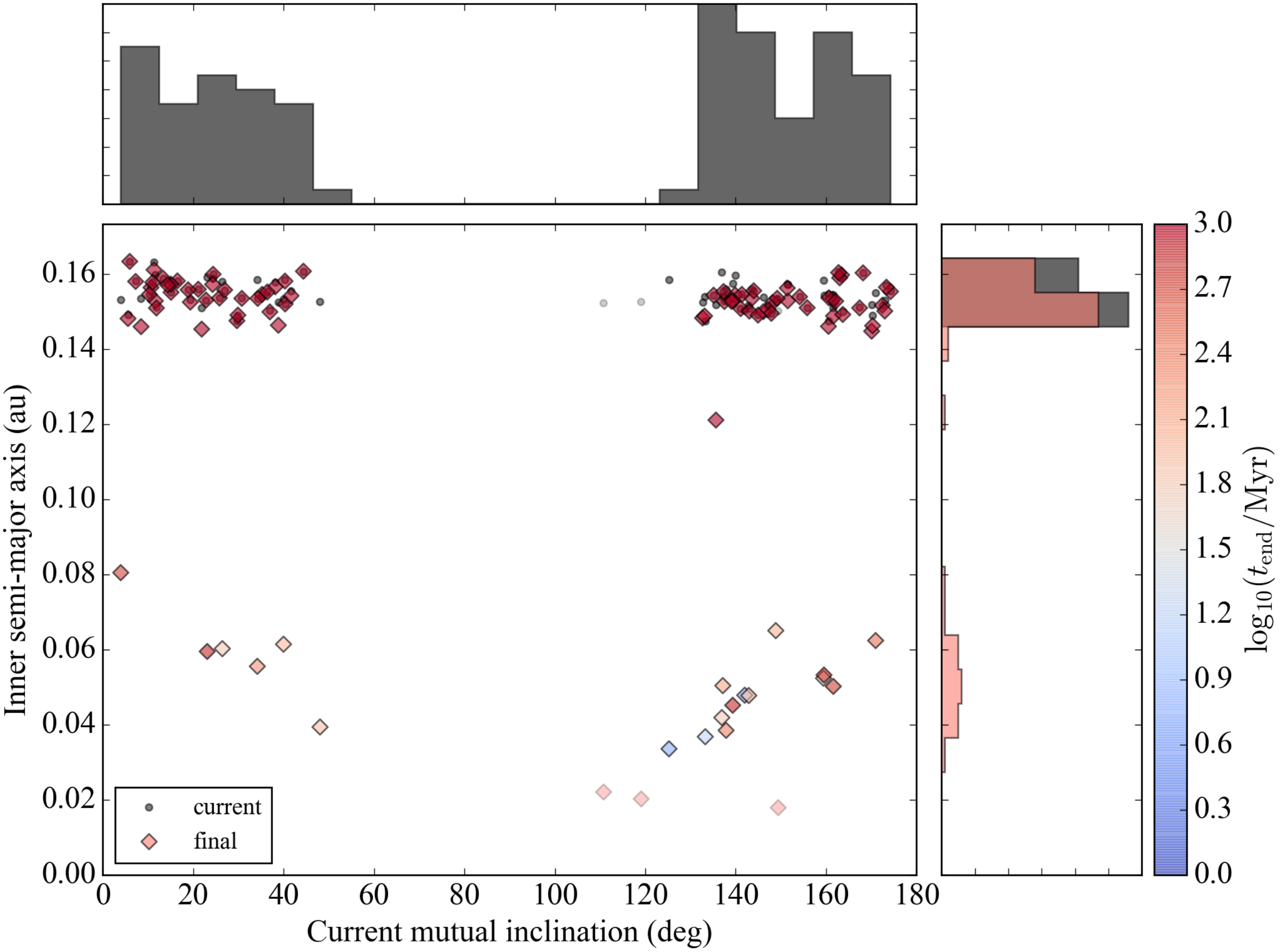}{0.49\textwidth}{(a) Kepler-448}
	\fig{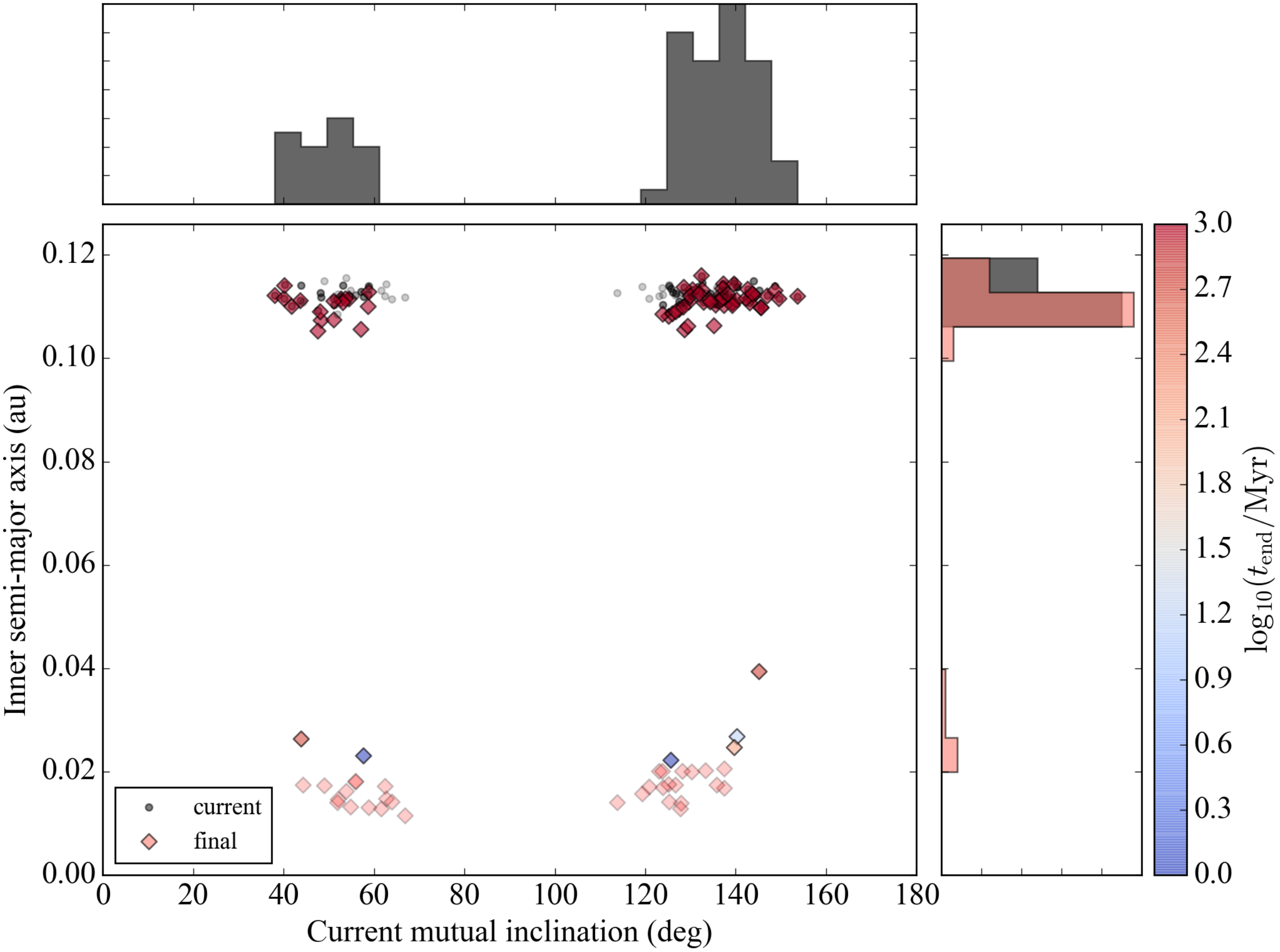}{0.49\textwidth}{(b) Kepler-693}
	\caption{Inner semi-major axes versus current mutual inclinations after $1\,\mathrm{Gyr}$ evolution in the presence of tidal dissipation with the planetary $t_{\rm V}=1.5\,\mathrm{yr}$ and stellar $t_{\rm V}=50\,\mathrm{yr}$ (diamonds and red histograms), along with their initial values (black circles and histograms). The color of each point shows the stopping time of the simulations, $t_{\rm end}$, defined as the shorter of $1\,\mathrm{Gyr}$ and the time when the inner orbit is circularized ($a_1<0.1\,\mathrm{au}$ and $e_1<0.01$). Models with minimum periastrons inside the Roche limit ($2.7\,R_\star$) are shown with transparent colors and not included in the histograms.}
	\label{fig:a1end}
\end{figure*}

Figure \ref{fig:a1end} compares the initial and final semi-major axes from those simulations. We see that some solutions do survive the tidal disruption and evolve into HJs within $1\,\mathrm{Gyr}$.
Such an outcome is rarer in the Kepler-693 system than in the Kepler-448 system. This is consistent with the expectation that the former system is likely older than the latter at least by a factor of a few. In fact, this kind of path may be even rarer than it seems in the right histograms of Figure \ref{fig:a1end}, because some of the survived HJs (shown with bluer colors) have experienced the circularization on a much shorter timescale compared to the system age due to the rapid eccentricity surge: if we are random observers in time, it is {\it a priori} unlikely to observe a system with such a short remaining lifetime compared to the current age \citep{1993Natur.363..315G}. However, I do not attempt to correct for this effect here, given that the outcome will be sensitive to the uncertain tidal parameter in any case. If correctly taken into account, this kind of argument will potentially allow for better constraints on the system parameters. 

\subsection{Implications for the In Situ Formation Scenario}\label{ssec:secular_insitu}

\red{
While the observed properties of the two inner planets are consistent with those of migrating WJs, presence of the outer (sub)stellar companions as close as $1.5\,\mathrm{au}$ challenges the high-eccentricity migration scenario from beyond the snow line. Planets in S-type orbits around tight binaries with periastron distances less than $10\,\mathrm{au}$ have also been reported around KOI-1257 \citep{2014A&A...571A..37S}, Kepler-444 \citep{2016ApJ...817...80D}, HD 59686 \citep{2016A&A...595A..55O}, and possibly $\nu$ Octantis \citep{2016MNRAS.460.3706R}. If confirmed to be a low-mass star, Kepler-693c has the smallest peristron among such stellar companions.
}

\red{
A similar issue has also been discussed for WJs with outer planetary-mass companions \citep{2016AJ....152..174A}: the outer orbits in these systems, if primordial, are in most cases too small for the inner WJs to have migrated from $\gtrsim1\,\mathrm{au}$. In addition, population synthesis simulations of high-eccentricity migration from $\gtrsim1\,\mathrm{au}$, either via the companion on a wide orbit \citep{2016MNRAS.456.3671A, 2016ApJ...829..132P} or secular chaos in multiple systems \citep{2017MNRAS.464..688H}, have difficulty in producing a sufficient number of WJs relative to HJs. These may also argue for the WJ formation via disk migration or in situ. Considering the prevalence of compact super-Earth systems revealed by {\it Kepler}, the latter can well be possible theoretically \citep{2014ApJ...797...95L} and may also have observational supports \citep{2016ApJ...825...98H}.
}

\red{
The companions discovered in the Kepler-448 and Kepler-693 systems may further argue for the in situ origin. 
Such low-mass stellar or brown-dwarf companions on au-scale orbits may be formed via fragmentation at a larger separation followed by the orbital decay due to dissipative dynamical interactions involving gas accretion and disks, which proceed in $\lesssim1\,\mathrm{Myr}$ 
\citep{2002MNRAS.336..705B, 2009MNRAS.392..413S, 2012MNRAS.419.3115B}. 
This implies that giant-planet formation and migration must have completed very quickly if they preceded those of the outer companion.
Alternatively, the companions may have arrived at the current orbit well after disk migration and disk dispersal via chaotic dissolution of an initial triple-star system 
or an impulsive encounter with a passing star \citep{2007A&A...467..347M, 2012A&A...544A..97M}.
While these scenarios are compatible with the eccentric and inclined outer orbit, they may suffer from the fine-tuning problem. In the former scenario, for example, a binary orbit typically shrinks only by a factor of a few, limited by energy conservation \citep{2007A&A...467..347M}. The outcomes are likely more diverse in the latter, but only a small fraction of them usually constitutes a suitable solution \citep{2012A&A...544A..97M}, and such close encounters as to alter the binary orbit significantly are likely rare when the planet formation is completed, even in a cluster with $\sim10^3$ stars \citep{2006ApJ...641..504A}. Also note that tidal friction associated with the close encounter with the primary \citep[e.g.,][]{1998MNRAS.300..292K} is unavailable to shrink the binary orbit in the presence of the inner planet.
Considering these possible difficulties of the alternative scenarios, in situ formation in a tight binary seems to be an attractive possibility that provides simple solutions both for our two systems and for other theoretical and observational issues, 
although the disk migration followed by rearrangement of the outer orbit cannot be excluded.}

\red{
This motivates us to examine whether the moderate eccentricities of our WJs can be explained in the in-situ framework, in which a near-circular orbit is normally expected.
Here we consider a specific form of question, in the same spirit as \citet{2017arXiv170600084A}:\footnote{See \citet{2017MNRAS.468.3000M} for other possible pathways of eccentric WJ formation.} suppose that the inner WJs were produced into circular orbits with the current semi-major axes, can their observed non-zero eccentricities be explained by the perturbation from the detected companions? To see this, I perform a similar set of simulations as in Section \ref{ssec:secular_ecc} setting $e_1=\omega_1=0$ initially, while sampling the other parameters from the posterior. Note that this experiment is applicable to the disk migration case as well, whose outcome is also a short-period planet on a circular orbit.}

\begin{figure*}
	\centering
	\fig{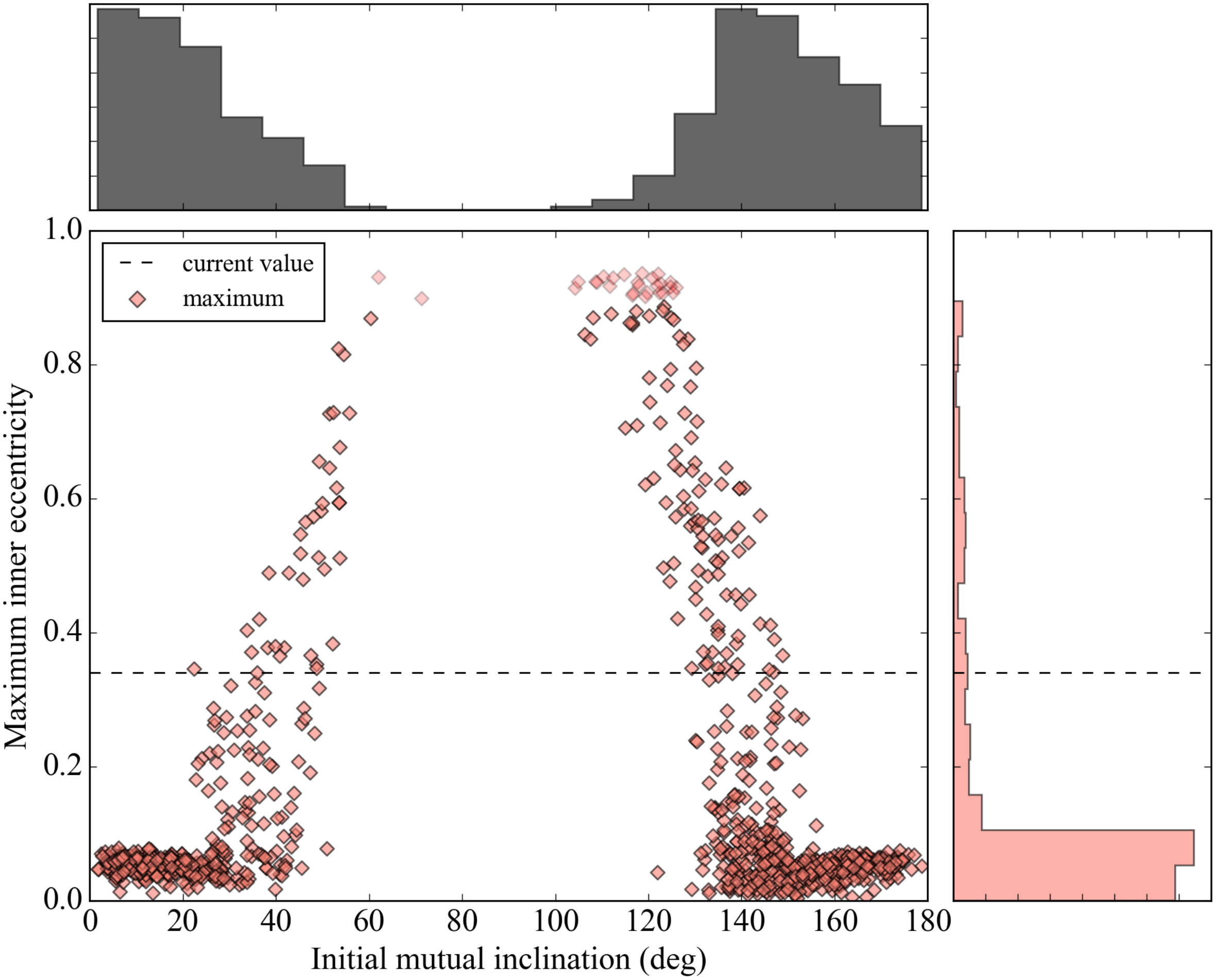}{0.49\textwidth}{(a) Kepler-448}
	\fig{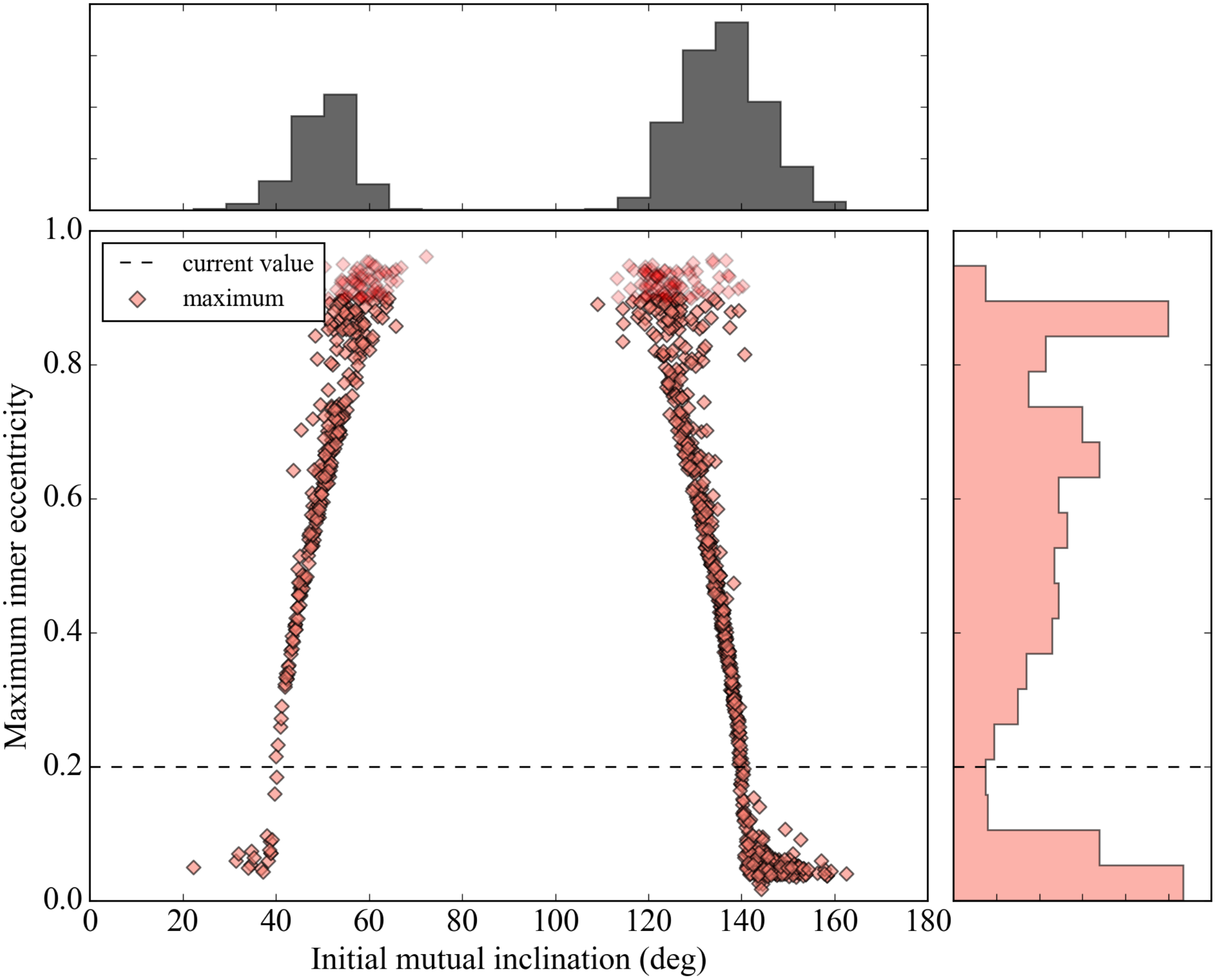}{0.49\textwidth}{(b) Kepler-693}
	\caption{Maximum eccentricities of the inner WJs achieved during secular evolution started from circular orbits against the current mutual orbital inclinations. The horizontal dashed lines show the median eccentricities from the dynamical modeling. Models with minimum periastrons inside the Roche limit are shown with transparent colors and not included in the histograms.}
	\label{fig:e1max}
\end{figure*}

Figure \ref{fig:e1max} summarizes the maximum inner eccentricities achieved during the $10\,\mathrm{Myr}$ evolution against the current mutual inclination. The maximum value exceeds the current best-fit value (horizontal dashed line) for $14\%$ and $76\%$ of the solutions that did not lead to tidal disruption in the Kepler-448 and Kepler-693 systems, respectively. Thus the current architecture can indeed be compatible with the initially circular orbits. The sequences of the maximum $e_1$ ($e_{\rm 1,max}$) and initial $i_{21}$ ($i_{\rm 21, init}$) roughly follows the relation $\sqrt{1-e_{\rm 1,max}^2}\sqrt{3/5}=\cos i_{\rm 21, init}$. 

Considering the arguments in Section \ref{ssec:secular_tau}, it is also conceivable that the inner WJs were formed into circular orbits at larger semi-major axes than observed now (but still inside the snow line) and have migrated to the current orbits via the tidal migration. In this case, excitation of eccentricity should have been easier because the gravitational interaction with the companion was initially stronger. This process might serve as yet another path of HJ formation: some HJs may have been isolated WJs formed in situ or via disk migration, whose orbits were shrunk due to the tidal high-eccentricity migration driven by a close companion.

\newpage

\section{Discussions}\label{sec:discussions}

\subsection{Follow-up Observations}\label{ssec:future}

While the future observations of transit times and durations will surely improve the constraint on the outer period and eccentricity, I did not find any systematic dependence of the future TTV and TDV evolutions on the current mutual orbital inclination at least within about 10 years. Moreover, it is challenging to observe a full transit from the ground, due to the long transit duration ($\simeq7\,\mathrm{hr}$) for Kepler-448b and faintness ($V=16.8$) of the host star for Kepler-693b.

How about RV observations? Based on the constraints from TTVs and TDVs, the RV semi-amplitude due to the outer companion is expected to be $2.7^{+0.7}_{-0.6}\times10^2\,\mathrm{m\,s^{-1}}$ for Kepler-448 and $2.7^{+0.9}_{-0.6}\,\mathrm{km\,s^{-1}}$ for Kepler-693. The variation may be observable for Kepler-693 with a large telescope, while the detection is implausible for Kepler-448 with $v\sin i_\star\simeq60\,\mathrm{km\,s^{-1}}$.

Instead, {\it Gaia} astrometry \citep{2001A&A...369..339P} is promising to detect the outer binary motion and better determine the mutual inclination. Since the misalignment in the sky plane $\Omega_{21}$ is dynamically well constrained in both systems (and the inner planets are transiting), inclinations of the outer orbits $i_2$, if measured, significantly improve the constraint on $i_{21}$. Based on the dynamical modeling, the expected astrometric displacements due to the outer companions are $1.4^{+1.2}_{-0.5}\times10^2\,\mu\mathrm{as}$ for Kepler-448 with $V=11.4$ \citep[assuming the distance $d=426\pm40\,\mathrm{pc}$ from][]{2015A&A...579A..55B} and $4.4^{+3.4}_{-1.5}\times10^2\,\mu\mathrm{as}$ for Kepler-693 with $V=16.8$ (assuming $d=1110\,\mathrm{pc}$ from the isochrone fit). They are both well above the astrometric precision expected after the nominal five-year mission \citep{2014ApJ...797...14P}.

\subsection{Frequency of Close and Massive Companions to WJs}

It is admittedly difficult to evaluate the completeness of our TTV search due to the complex dependence of the signal on the system parameters. Nevertheless, our detection of the close and massive companions in two systems, among the sample of $23$ WJs, suggests such an architecture is not extremely uncommon. We also need to take into account that the two systems would not have been detected if the periastron passage did not occur during the {\it Kepler} mission. The ratios of the outer orbital periods ($P_2\sim2500\,\days$ for Kepler-448c and $P_2\sim1800\,\days$ for Kepler-693c) to the {\it Kepler} observing duration ($\sim1400\,\mathrm{days}$) suggest that there may be one or two more WJs with similar companions in the {\it Kepler} sample that have evaded our search. This crude estimate seems compatible with the theoretical argument by \citet{2016ApJ...829..132P} that roughly $20\%$ of WJs can be accounted for by high-eccentricity migration, \red{although the observed architectures of our systems may not support the migration from $\gtrsim1\,\mathrm{au}$ via this process as assumed in \citet{2016ApJ...829..132P}.}

\section{Summary and Conclusion}\label{sec:summary}

This paper reported the discovery of close companions to two transiting WJs via their TTVs and TDVs. The companions have masses comparable to a brown dwarf or a low-mass star ($22^{+7}_{-5}\,M_{\rm Jup}$ and $150^{+60}_{-40}\,M_{\rm Jup}$), and they are on highly eccentric orbits ($e\gtrsim0.5$) with small periastron distances ($\simeq1.5\,\mathrm{au}$). For the companion of Kepler-693b, a large mutual orbital inclination ($\simeq50\degr$) with respect to the inner planetary orbit is indicated by TDVs, while the constraint on the mutual inclination is weak for the Kepler-448 system. They are among the few systems with constraints on mutual inclinations, 
and that inferred for the Kepler-693 system is the largest ever determined dynamically for planetary systems. The value is indeed large enough for the eccentricity oscillation via the Kozai mechanism to occur: more than $90\%$ of the solutions inferred for Kepler-693b (and some $10\%$ for Kepler-448b) imply that the inner WJs' eccentricities exhibit a large enough oscillations for tidal dissipation to affect the inner orbits significantly, by bringing $a(1-e^2)$ less than $0.1\,\mathrm{au}$. The corresponding migration timescales can be compatible with the hypothesis that the inner WJs are tidally migrating to evolve into HJs, for a wide range of viscous timescales. 

The architectures of the two systems support the scenario that eccentric WJs are currently undergoing eccentricity oscillation induced by a close companion and experiencing the slow tidal migration where the orbit shrinks only at the high-eccentricity phase. On the other hand, the origin of the current highly eccentric/inclined configuration is still unclear. Specifically, they may not fit into the classical picture that close-in gas giants have migrated from {\it beyond} the snow line, given the close and (sub-)stellar nature of the outer companions. 
Formation of gas giants within the snow line onto circular orbits, followed by eccentricity excitation by the companion, therefore seems another viable option to be pursued,
\red{although the companion may instead have arrived at the current orbit after disk migration of the inner WJ. Regardless of the origin of the current configuration, the long-term evolution simulation demonstrates a new pathway of HJ formation via ``high-eccentricity" migration of a WJ formed in situ or via disk migration.
}

\acknowledgements

The author is grateful to the entire {\it Kepler} team for making the revolutionary data available. The author thanks Hajime Kawahara, Yasushi Suto, Yuxin Xue, Saul Rappaport, Josh Winn, Sho Uehara, Sarah Ballard, Gongjie Li, Tim Morton, Dong Lai, Kaloyan Penev, Adrian Price-Whelan, Chelsea Huang, George Zhou, Bekki Dawson, Cristobal Petrovich, Amaury Triaud, Kengo Tomida, Kaitlin Kratter, Eve Ostriker, and Ji-Ming Shi for helpful conversations and insightful discussions; and Tam\'as Borkovits for answering questions on the analytic ETV formula.
Comments on the manuscript from Dong Lai, Josh Winn, and an anonymous referee 
were also very helpful in improving the paper and enriching its content.
K.M. acknowledges the support by JSPS Research Fellowships for Young Scientists (No. 26-7182) and also by the Leading Graduate Course for Frontiers of Mathematical Sciences and Physics. This work was performed in part under contract with the California Institute of Technology (Caltech)/Jet Propulsion Laboratory (JPL) funded by NASA through the Sagan Fellowship Program executed by the NASA Exoplanet Science Institute.

This paper includes data collected by the {\it Kepler} mission. Funding for the {\it Kepler} mission is provided by the NASA Science Mission directorate. Some of the data presented in this paper were obtained from the Mikulski Archive for Space Telescopes (MAST). STScI is operated by the Association of Universities for Research in Astronomy, Inc., under NASA contract NAS5-26555. Support for MAST for non-HST data is provided by the NASA Office of Space Science via grant NNX09AF08G and by other grants and contracts. This research has made use of the NASA Exoplanet Archive, which is operated by the California Institute of Technology, under contract with the National Aeronautics and Space Administration under the Exoplanet Exploration Program. This research has made use of NASA's Astrophysics Data System Bibliographic Services. 

\appendix

\section{Analytic TTV Formula for Hierarchical Triple Systems}\label{app:formula}

In a part of the TTV analysis, we adopt the analytic timing formula for hierarchical three-body systems 
by \citet{2015MNRAS.448..946B},
which takes into account the eccentricities of both inner and outer orbits
and arbitrary mutual inclination between them
\citep[see also][for its applications]
{2011A&A...528A..53B, 2015MNRAS.448..946B, 2016MNRAS.455.4136B}.
Among various effects that could possibly affect the observed transit times,
we include two of the most important effects:
light-travel time effect (LTTE) and $P_2$-timescale dynamical effects up to
the quadrupole order.
The other effects including the octupole-level dynamical effects
and short-term perturbations are at least an order-of-magnitude smaller
than the quadrupole terms and thus neglected 
\citep[see][for a complete discussion of these effects]{2015MNRAS.448..946B}.
Note that, in this appendix, $+Z$-axis is taken {\it away} from the observer's direction; this definition is opposite to the one in the main text, and so arguments of periastron in the formulae below differ by $\pi$ from the ones in the main text.

\subsection{Formula}

We model the timing variations from the linear ephemeris due the LTTE effect and the $P_2$-timescale dynamical effect. The LTTE term is given by
\begin{equation}
	\label{eq:ltte}
	\Delta_{\rm LTTE}
	=-\frac{a_2\sin i_2}{c}\frac{m_\mathrm{c}}{m_\star+m_\mathrm{b}+m_\mathrm{c}} \frac{(1-e_2^2)\sin u_2}{1+e_2\cos v_2},
\end{equation}
where $v_2$ is the true anomaly and $u_2\equiv\omega_2+v_2$ is the true longitude. The $P_2$-timescale dynamical effect is modeled up to the quadrupole order as follows \citep[][equation 5]{2015MNRAS.448..946B}:
\begin{equation}
	\label{eq:quad}
	\Delta_1
	=\frac{P_1}{2\pi} A_{\rm L1}(1-e_1^2)^{1/2}\left[\delta_{\rm tidal}+\delta_{\rm ecc1}+\delta_{\rm ecc2}+\delta_{\rm noncopl}\right].
\end{equation}
Here the overall amplitude is fixed by the factor
\begin{equation}
	\label{eq:quad_amp}
	A_{\rm L1}=\frac{15}{8}\frac{m_\mathrm{c}}{m_\star+m_\mathrm{b}+m_\mathrm{c}}\frac{P_1}{P_2} \,(1-e_2^2)^{-3/2},
\end{equation}
and each TTV component is given by
\begin{equation}
	\label{eq:quad_tidal}
	\delta_{\rm tidal}=\left[\frac{8}{15}f_1+\frac{4}{5}K_1\right]\mathcal{M},
\end{equation}
\begin{equation}
	\label{eq:quad_ecc1}
	\delta_{\rm ecc1}=(1+\cos i_{21})\left\{K_{11}\mathcal{S}\left[2u_2-2(n_2-n_1)\right]-K_{12}\mathcal{C}\left[2u_2-2(n_2-n_1)\right]\right\},
\end{equation}
\begin{equation}
	\label{eq:quad_ecc2}
	\delta_{\rm ecc2}=(1-\cos i_{21})\left\{K_{11}\mathcal{S}\left[2u_2-2(n_2+n_1)\right]+K_{12}\mathcal{C}\left[2u_2-2(n_2+n_1)\right]\right\},
\end{equation}
\begin{equation}
	\label{eq:quad_noncopl}
	\delta_{\rm noncopl}=\sin^2 i_{21}\left(K_{11}\cos2n_1+K_{12}\sin2n_1-\frac{2}{5}f_1-\frac{3}{5}K_1\right)\left[2\mathcal{M}-\mathcal{S}(2u_2-2n_2)\right],
\end{equation}
where
\begin{equation}
	\label{eq:mfunc}
	\mathcal{M}=v_2-l_2+e_2\sin v_2,
\end{equation}
\begin{equation}
	\label{eq:sfunc}
	\mathcal{S}(2u_2)=\sin2u_2+e_2\left[\sin(u_2+\omega_2)+\frac{1}{3}\sin(3u_2-\omega_2)\right],
\end{equation}
\begin{equation}
	\label{eq:cfunc}
	\mathcal{C}(2u_2)=\cos2u_2+e_2\left[\cos(u_2+\omega_2)+\frac{1}{3}\cos(3u_2-\omega_2)\right],
\end{equation}
\begin{equation}
	f_1=1+\frac{25}{8}e_1^2+\frac{15}{8}e_1^4+\frac{95}{64}e_1^6+\mathcal{O}(e_1^8),
\end{equation}
\begin{align}
	\notag
	K_1(e_1, \omega_1)
	=&-e_1\sin\omega_1+\left(\frac{3}{4}e_1^2+\frac{1}{8}e_1^4+\frac{3}{64}e_1^6\right)\cos2\omega_1+\left(\frac{1}{2}e_1^3+\frac{3}{16}e_1^5\right)\sin3\omega_1\\
	&-\left(\frac{5}{16}e_1^4+\frac{3}{16}e_1^6\right)\cos4\omega_1-\frac{3}{16}e_1^5\sin5\omega_1+\frac{7}{64}e_1^6\cos6\omega_1+\mathcal{O}(e_1^7),
\end{align}
\begin{align}
	\notag
	K_{11}(e_1, \omega_1)
	=&\frac{3}{4}e_1^2+\frac{3}{16}e_1^4+\frac{3}{32}e_1^6
	+\left(e_1+\frac{1}{2}e_1^3+\frac{1}{4}e_1^5\right)\sin\omega_1
	+\left(\frac{51}{40}e_1^2+\frac{37}{80}e_1^4+\frac{241}{640}e_1^6\right)\cos2\omega_1\\
	&-\frac{3}{16}e_1^3\sin3\omega_1-\left(\frac{1}{16}e_1^4-\frac{1}{16}e_1^6\right)\cos4\omega_1
	-\frac{1}{16}e_1^5\sin5\omega_1+\frac{3}{64}e_1^6\cos6\omega_1
	+\mathcal{O}(e_1^7),
\end{align}
\begin{align}
	\notag
	K_{12}(e_1, \omega_1)=
	&-\left(e_1-\frac{1}{2}e_1^3-\frac{1}{4}e_1^5\right)\cos\omega_1
	+\left(\frac{51}{40}e_1^2+\frac{87}{80}e_1^4+\frac{541}{640}e_1^6\right)\sin2\omega_1\\
	&-\frac{3}{16}e_1^3\cos3\omega_1-\left(\frac{1}{16}e_1^4+\frac{5}{32}e_1^6\right)\sin4\omega_1
	+\frac{1}{16}e_1^5\cos5\omega_1+\frac{3}{64}e_1^6\sin6\omega_1
	+\mathcal{O}(e_1^7).
\end{align}
The angles $n_1$ and $n_2$ are the directions of $Z>0$ part of the line of intersection of the inner and outer orbits measured from the ascending nodes of the two orbits, defined as in figure 1 of \citet{2015MNRAS.448..946B} between $[0, \pi]$, and $l_2$ is the mean anomaly.
Note that $K_{11}(e_1, \pi-\omega_1)=K_{11}(e_1, \omega_1)$ and $K_{12}(e_1, \pi-\omega_1)=-K_{12}(e_1, \omega_1)$.

\subsection{Conversion of the Angles} \label{app:conversion}

In the main body of the paper, we did not use the physically most natural parametrization of the angles because it complicates the assignment of the prior in the {\tt MultiNest} fitting. Here we summarize how to relate our set of fitted angles  ($i_1$, $i_2$, $\Omega_{21}$) to that of physical angles ($i_{21}$, $n_1$, $n_2$) in the analytic formula by \citet{2015MNRAS.448..946B}, since this case is not covered in their Appendix D.

The mutual inclination $i_{21}$ can be computed as usual:
\begin{equation}
	\label{eq:i21}
	\cos i_{21}
	= \cos i_1 \cos i_2 + \sin i_1 \sin i_2 \cos\Omega_{21}.
\end{equation}
Note that $\sin i_{21}\neq0$ for $\Omega_{21}\neq0$
since $(i_1, i_2)\neq(0, 0)$ nor $(\pi, \pi)$ for transiting systems 
as discussed in this paper.
For $\cos\Omega_{21}=\pm 1$, the above equation reduces to
\begin{equation}
	\cos i_{21}=\cos(i_2\mp i_1).
\end{equation}

Let us first consider the case when $\sin i_{21}\neq0$.
If $\sin\Omega_{21}\neq0$,
the sine and cosine rules of the spherical trigonometry yield
\begin{equation}
	\frac{\sin i_{21}}{|\sin \Omega_{21}|}
	= \frac{\sin i_2}{\sin n_1} = \frac{\sin i_1}{\sin n_2}
\end{equation}
and
\begin{equation}
	\cos n_1 = \cos \Omega_{21} \cos n_2 + |\sin \Omega_{21}| \sin n_2 \cos i_2,
\end{equation}
\begin{equation}
	\cos n_2 = \cos \Omega_{21} \cos n_1 + |\sin \Omega_{21}| \sin n_1 \cos (\pi-i_1),
\end{equation}
for $\sin\Omega_{21}<0$ case.
For $\sin\Omega_{21}>0$, the cosine rules are replaced by
\begin{equation}
	\cos n_1 = \cos \Omega_{21} \cos n_2 + \sin \Omega_{21} \sin n_2 \cos (\pi-i_2),
\end{equation}
\begin{equation}
	\cos n_2 = \cos \Omega_{21} \cos n_1 + \sin \Omega_{21} \sin n_1 \cos i_1.
\end{equation}
In fact, the two cases can be written in a single form as 
\begin{equation}
	\cos n_1 = \cos \Omega_{21} \cos n_2 - \sin \Omega_{21} \sin n_2 \cos i_2,
\end{equation}
\begin{equation}
	\cos n_2 = \cos \Omega_{21} \cos n_1 + \sin \Omega_{21} \sin n_1 \cos i_1.
\end{equation}
Thus, for a non-zero mutual inclination we have
\begin{equation}
	\sin n_{1,2} = \mathrm{sgn}(\sin\Omega_{21})\,
	\frac{\sin i_{2,1}}{\sin i_{21}} \sin\Omega_{21}
	= \frac{\sin i_{2,1}}{\sin i_{21}} |\sin\Omega_{21}|
\end{equation}
and 
\begin{equation}
	\cos n_1
	= \frac{\mathrm{sgn}(\sin\Omega_{21})}{\sin i_{21}}
	\left( - \sin i_1\cos i_2 + \cos\Omega_{21}\cos i_1\sin i_2 \right),
\end{equation}
\begin{equation}
	\cos n_2
	= \frac{\mathrm{sgn}(\sin\Omega_{21})}{\sin i_{21}}
	\left( \cos i_1\sin i_2 - \cos\Omega_{21}\sin i_1\cos i_2 \right).
\end{equation}

If $\sin i_{21}=0$, we may just define $n_2-n_1=0$ for $i_{21}=0$
and $n_2+n_1=\pi$ for $i_{21}=\pi$ to correctly compute the non-vanishing 
terms of TTVs.
Other than $i_1=i_2=0$ or $\pi$, this includes the two cases:
(i) $i_1-i_2=0$ and $\Omega_{21}=0$
and (ii) $i_1+i_2=\pi$ and $\Omega_{21}=\pi$.
As shown in Table \ref{tab:limit}, we have $n_1-n_2=0$ for either $i_2-i_1\to\pm0$ in case (i).
Similarly in case (ii), we have $n_1+n_2=\pi$ for either $i_2+i_1-\pi\to\pm0$.
Although the other combination is ambiguous, it appears only 
in the vanishing terms of the TTV formula.

\begin{table}[h]
	\centering	
	\caption{$\cos n_{1,2}$ for $\sin\Omega_{21}\to0$}
	\label{tab:limit}
	\begin{tabular}{ccc}
	\hline\hline
		&	$(\cos n_1, \cos n_2)$	&	$(n_1, n_2)$\\
	\hline
	$\cos\Omega_{21}\to1$	
		&	$\mathrm{sgn}(\sin\Omega_{21})\left(\frac{\sin(i_2-i_1)}{\sin i_{21}}, \frac{\sin(i_2-i_1)}{\sin i_{21}}\right)$
		& $(0, 0)$ or $(\pi, \pi)$ for $i_2-i_1\to0$\\
	$\cos\Omega_{21}\to-1$
		&	$\mathrm{sgn}(\sin\Omega_{21})\left(-\frac{\sin(i_2+i_1)}{\sin i_{21}}, \frac{\sin(i_2+i_1)}{\sin i_{21}}\right)$
		& $(0, \pi)$ or $(\pi, 0)$ for $i_2+i_1\to\pi$\\
	\hline \hline
	\end{tabular}
\end{table}

\section{Results of Analytic and Numerical TTV Analyses} \label{app:ttv}

\subsection{Comparison of the Two Results}

For TTVs, I performed both an analytic fit with a wider prior range and a numerical fit with a narrower prior range. As shown in Tables \ref{tab:ttv_koi12} and \ref{tab:ttv_koi824}, I found consistent posteriors from the two analyses; this agreement validates our numerical procedure. The best-fit models are basically indistinguishable from those in Figures \ref{fig:bestfit_koi12} and \ref{fig:bestfit_koi824}.

\begin{deluxetable*}{lcccc}
	\tablewidth{0pt}
	\tablecolumns{5}
	\tablecaption{Parameters of the Kepler-448 System from the Analytical and Numerical TTV Analyses \label{tab:ttv_koi12}}
	\tablehead{
		& \multicolumn{2}{c}{Analytic} & \multicolumn{2}{c}{Numerical}\\
		\colhead{Parameter} & \colhead{Posterior} & \colhead{Prior} & \colhead{Posterior}  & \colhead{Prior}
	}
	\startdata
	{\bf Fitted Parameters}\\
	({\it Inner Orbit})\\
	1. Time of inferior conjunction 
		& $128.7414^{(+7)}_{(-3)}$ 
		& $\uni(128.7, 128.8)$
		& $128.7418^{(+2)}_{(-2)}$
		& $\uni(128.73, 128.75)$\\
		\quad $t_{\rm ic, 1}$ ($\kbjd$) & & &\\
	2. Orbital period $P_1$ (day)
		& $17.855219^{(+6)}_{(-13)}$ 
		& $\luni(17.85, 17.86)$
		& $17.855179^{(+7)}_{(-6)}$		
		& $\luni(17.8551, 17.8553)$\\
	3. Orbital eccentricity $e_1$
		& $0.5^{+0.3}_{-0.2}$ 
		& $\uni(0, 0.95)$
		& $0.5^{+0.1}_{-0.2}$ 
		& $\uni(0, 0.7)$\\
	4. Argument of periastron $\omega_1$ (deg)
		& $-43^{+158}_{-104}$ & $\uni(-180, 180)$
		& $-15^{+166}_{-140}$ 
		& $\uni(-180, 180)$\\
	5. Cosine of orbital inclination $\cos i_1$
		& $0$ (fixed) & \nodata & $0$ (fixed) & \nodata\\
	({\it Outer Orbit})\\
	6. Time of the periastron passage
		& $1076^{+10}_{-10}$ & $\uni(120, 1600)$ 
		& $1072^{+12}_{-10}$ & $\uni(900, 1200)$\\
		\quad $\tau_2$ ($\kbjd$) &&&\\
	7. Periastron distance over 
		& $9.3^{+0.4}_{-0.3}$ & $\luni(1, 50)$ 
		& $9.4^{+0.4}_{-0.4}$ & $\luni(5, 20)$\\
		\quad inner semi-major axis $q_2/a_1$&&&\\
	8. Orbital eccentricity $e_2$
		& $0.54^{+0.10}_{-0.05}$ & $\uni(0, 0.95)$
		& $0.60^{+0.12}_{-0.07}$ & $\uni(0, 0.95)$\\
	9. Argument of periastron $\omega_2$ (deg)
		& $-91^{+52}_{-29}$ &  $\uni(-180, 180)$ 
		& $-116^{+58}_{-34}$ & $\uni(-180, 180)$\\
	10. Cosine of orbital inclination $\cos i_2$
		& $0.1^{+0.7}_{-0.8}$ & $\uni(-1, 1)$ 
		& $0.1^{+0.6}_{-0.7}$ & $\uni(-1, 1)$\\
	11. Relative longitude of 
		& $-5^{+162}_{-155}$ & $\uni(-180, 180)$
		& $1^{+135}_{-131}$ & $\uni(-180, 180)$\\
		\quad ascending node $\Omega_{21}$ (deg)\tablenotemark{a} &&&\\
	({\it Physical Properties})\\
	12. Mass of Kepler-448 $m_\star$ ($M_\odot$)
		& $1.54^{+0.11}_{-0.09}$	& $\gaus(1.51, 0.09, 0.14)$
		& $1.54^{+0.11}_{-0.09}$ 	& $\gaus(1.51, 0.09, 0.14)$\\
	13. Mass of Kepler-448b $m_\mathrm{b}$ ($M_{\rm Jup}$)
		& $1$ (fixed) & \nodata & $1$ (fixed)
		& \nodata\\
	14. Mass of Kepler-448c $m_\mathrm{c}$ ($M_{\rm Jup}$)
		& $12^{+10}_{-4}$ & $\luni(10^{-4}M_\odot, 0.3M_\odot)$
		& $15^{+7}_{-4}$ & $\luni(10^{-3}M_\odot, 0.1M_\odot)$\\
	({\it Jitters})\\
	15. Transit time jitter $\sigma_{\rm TTV}$ ($10^{-4}\,\mathrm{day}$)
		& $2.2\pm0.3$ & $\luni(0.05, 5)$
		& $2.2\pm0.3$ & $\luni(0.05, 5)$\\
	\\
	{\bf Derived Parameters}\\
	Outer orbital period $P_2$ (day)
		& $(1.6^{+0.8}_{-0.2})\times10^3$ & \nodata
		& $(2.0^{+1.4}_{-0.4})\times10^3$ & \nodata\\
	Inner semi-major axis $a_1$ (au)
		&  $0.154(3)$ & \nodata & $0.154(3)$
		& \nodata\\
	Outer semi-major axis $a_2$ (au)
		&  $3.1^{+0.9}_{-0.3}$ & \nodata & $3.6^{+1.5}_{-0.5}$ & \nodata\\
	Periastron distance of 
		&  $1.44^{+0.07}_{-0.06}$ & \nodata &  $1.45^{+0.07}_{-0.06}$ 
		& \nodata \\
		\quad the outer orbit $a_2(1-e_2)$ (au) &&&\\
	Mutual orbital inclination $i_{21}$ (deg)
		& $118^{+31}_{-68}$	& \nodata & $54^{+81}_{-28}$ & \nodata\\
	\enddata
	\tablecomments{The elements of the inner and outer orbits listed here
	are Jacobian osculating elements defined at the epoch 
	$\mathrm{BJD}=2454833+120$.
	The quoted values in the `Solution' columns are the median
	and $68\%$ credible interval of the marginal posteriors.
	Parentheses after values denote uncertainties in the last digit.
	The `combined' column shows the values from the marginal posterior 
	combining the two solutions; no value is shown if the combined 
	marginal posterior is multimodal. In the prior column,
	$\uni(a, b)$ and $\luni(a, b)$ denote the (log-)uniform priors between $a$ and $b$,
	$f(x)=1/(b-a)$ and $f(x)=x^{-1}/(\ln b - \ln a)$, respectively;
	$\gaus(a, b, c)$ means the asymmetric Gaussian prior 
	with the central value $a$ and lower and upper widths $b$ and $c$.}
	\tablenotetext{a}{Referenced to the ascending node of the inner orbit,
	whose direction is arbitrary.}
\end{deluxetable*}

\begin{deluxetable*}{lcccc}
	\tablewidth{0pt}
	\tablecolumns{5}
	\tablecaption{Parameters of the Kepler-693 System from the Analytical and Numerical TTV Analyses \label{tab:ttv_koi824}}
	\tablehead{
		& \multicolumn{2}{c}{Analytic} & \multicolumn{2}{c}{Numerical}\\
		\colhead{Parameter} & \colhead{Posterior} & \colhead{Prior} & \colhead{Posterior}  & \colhead{Prior}
	}
	\startdata
	{\bf Fitted Parameters}\\
	({\it Inner Orbit})\\
	1. Time of inferior conjunction 
		& $173.611^{(+2)}_{(-1)}$
		& $\uni(173.55, 173.65)$ 
		& $173.609^{(+1)}_{(-2)}$ 
		& $\uni(173.58, 173.64)$\\
		\quad $t_{\rm ic, 1}$ ($\kbjd$) & & &\\
	2. Orbital period $P_1$ (day)
		& $15.37565^{(+3)}_{(-7)}$ 
		& $\luni(15.37, 15.38)$
		& $15.37549^{(+11)}_{(-7)}$ 
		& $\luni(15.375, 15.376)$\\
	3. Orbital eccentricity $e_1$
		& $0.6\pm0.2$ 
		& $\uni(0, 0.95)$
		& $0.5\pm0.1$ 
		& $\uni(0, 0.7)$\\
	4. Argument of periastron $\omega_1$ (deg)
		& $64^{+72}_{-158}$
		& $\uni(-180, 180)$ 
		& $84^{+62}_{-93}$
		& $\uni(-180, 180)$\\
	5. Cosine of orbital inclination $\cos i_1$
		& $0$ (fixed) & \nodata & $0$ (fixed) & \nodata\\
	({\it Outer Orbit})\\
	6. Time of the periastron passage
		& $636^{+28}_{-16}$ 
		& $\uni(170, 1560)$
		& $660^{+23}_{-20}$ 
		& $\uni(500, 900)$\\
		\quad $\tau_2$ ($\kbjd$) &&&\\
	7. Periastron distance over 
		& $13^{+2}_{-2}$ 
		& $\luni(1, 50)$
		& $15^{+2}_{-2}$ 
		& $\luni(6, 25)$\\
		\quad inner semi-major axis $q_2/a_1$&&&\\
	8. Orbital eccentricity $e_2$
		& $0.50^{+0.18}_{-0.08}$ 
		& $\uni(0, 0.95)$
		& $0.7^{+0.1}_{-0.1}$ 		
		& $\uni(0, 0.95)$\\
	9. Argument of periastron $\omega_2$ (deg)
		& $60^{+58}_{-62}$ 
		& $\uni(-180, 180)$
		& $86^{+42}_{-47}$  
		& $\uni(-180, 180)$\\
	10. Cosine of orbital inclination $\cos i_2$
		& $0.0\pm0.3$ 
		& $\uni(-1, 1)$
		& $-0.1\pm0.4$ 
		& $\uni(-1, 1)$\\
	11. Relative longitude of 
		& $-6^{+167}_{-157}$ 
		& $\uni(-180, 180)$
		& $-21^{+178}_{-139}$  
		& $\uni(-180, 180)$\\
		\quad ascending node $\Omega_{21}$ (deg)\tablenotemark{a} &&&\\
	({\it Physical Properties})\\
	12. Mass of Kepler-693 $m_\star$ ($M_\odot$)
		& $0.80\pm0.03$	
		& $\gaus(0.79, 0.03, 0.05)$
		& $0.80\pm0.03$ 
		& $\gaus(0.79, 0.03, 0.05)$\\
	13. Mass of Kepler-693b $m_\mathrm{b}$ ($M_{\rm Jup}$)
		& $1$ (fixed) & \nodata & $1$ (fixed) & \nodata\\
	14. Mass of Kepler-693c $m_\mathrm{c}$ ($M_{\rm Jup}$)
		& $57^{+41}_{-21}$ 
		& $\luni(10^{-4}M_\odot, 0.3M_\odot)$
		& $94^{+37}_{-23}$  
		& $\luni(10^{-3}M_\odot, 0.3M_\odot)$\\
	({\it Jitters})\\
	15. Transit time jitter $\sigma_{\rm TTV}$ ($10^{-4}\,\mathrm{day}$)
		& $6^{+4}_{-3}$	
		& $\luni(0.05, 50)$
		& $6^{+4}_{-3}$ 
		& $\luni(0.05, 50)$\\
	\\
	{\bf Derived Parameters}\\
	Outer orbital period $P_2$ (day)
		& $(2.0^{+2.5}_{-0.6})\times10^3$ 
		& \nodata
		& $(5.3^{+8.0}_{-2.4})\times10^3$  
		& \nodata\\
	Inner semi-major axis $a_1$ (au)
		&  $0.112^{(+2)}_{(-1)}$
		& \nodata 
		& $0.112^{(+2)}_{(-1)}$ 
		& \nodata\\
	Outer semi-major axis $a_2$ (au)
	        & $3.0^{+2.1}_{-0.6}$
                & \nodata
                & $5.8^{+4.9}_{-1.9}$ 
		& \nodata\\
	Periastron distance of 
          	& $1.5\pm0.2$
		& \nodata
		& $1.6\pm0.2$ 
		& \nodata\\
		\quad the outer orbit $a_2(1-e_2)$ (au) &&&\\
	Mutual orbital inclination $i_{21}$ (deg)
		& $133^{+30}_{-115}$	 
		& \nodata
		& $133^{+30}_{-105}$	
		& \nodata\\
	\enddata
	\tablecomments{The elements of the inner and outer orbits listed here
	are Jacobian osculating elements defined at the epoch 
	$\mathrm{BJD}=2454833+170$.
	The quoted values in the `Solution' columns are the median
	and $68\%$ credible interval of the marginal posteriors.
	Parentheses after values denote uncertainties in the last digit.
	The `combined' column shows the values from the marginal posterior 
	combining the two solutions; no value is shown if the combined 
	marginal posterior is multimodal. In the prior column,
	$\uni(a, b)$ and $\luni(a, b)$ denote the (log-)uniform priors between $a$ and $b$,
	$f(x)=1/(b-a)$ and $f(x)=x^{-1}/(\ln b - \ln a)$, respectively;
	$\gaus(a, b, c)$ means the asymmetric Gaussian prior 
	with the central value $a$ and lower and upper widths $b$ and $c$.}
	\tablenotetext{a}{Referenced to the ascending node of the inner orbit,
	whose direction is arbitrary.}
\end{deluxetable*}

\subsection{Decomposition of the TTV Solutions Using the Analytic Formula}

\begin{figure*}
	\centering
	\fig{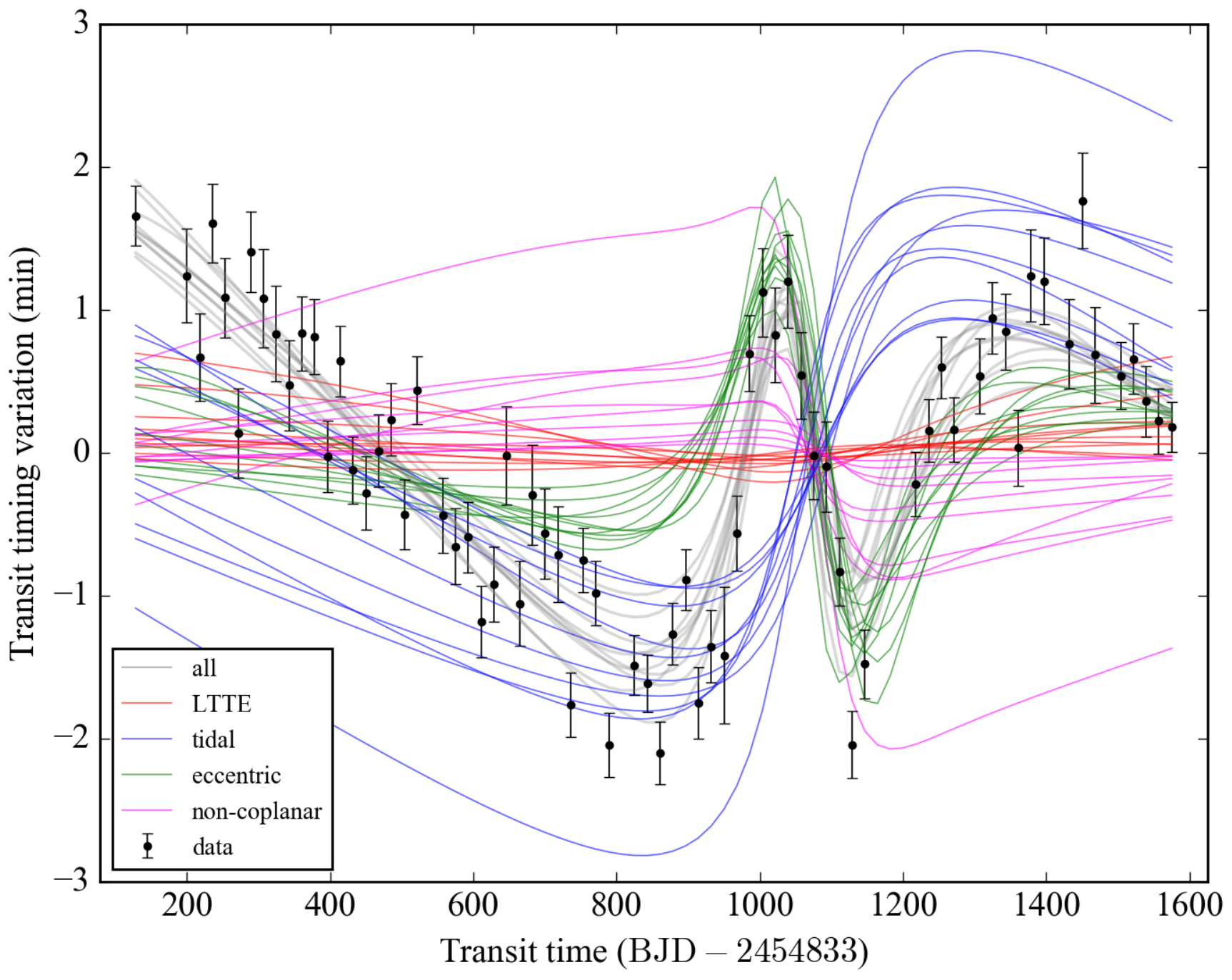}{0.49\textwidth}{(a) Kepler-448b}
	\fig{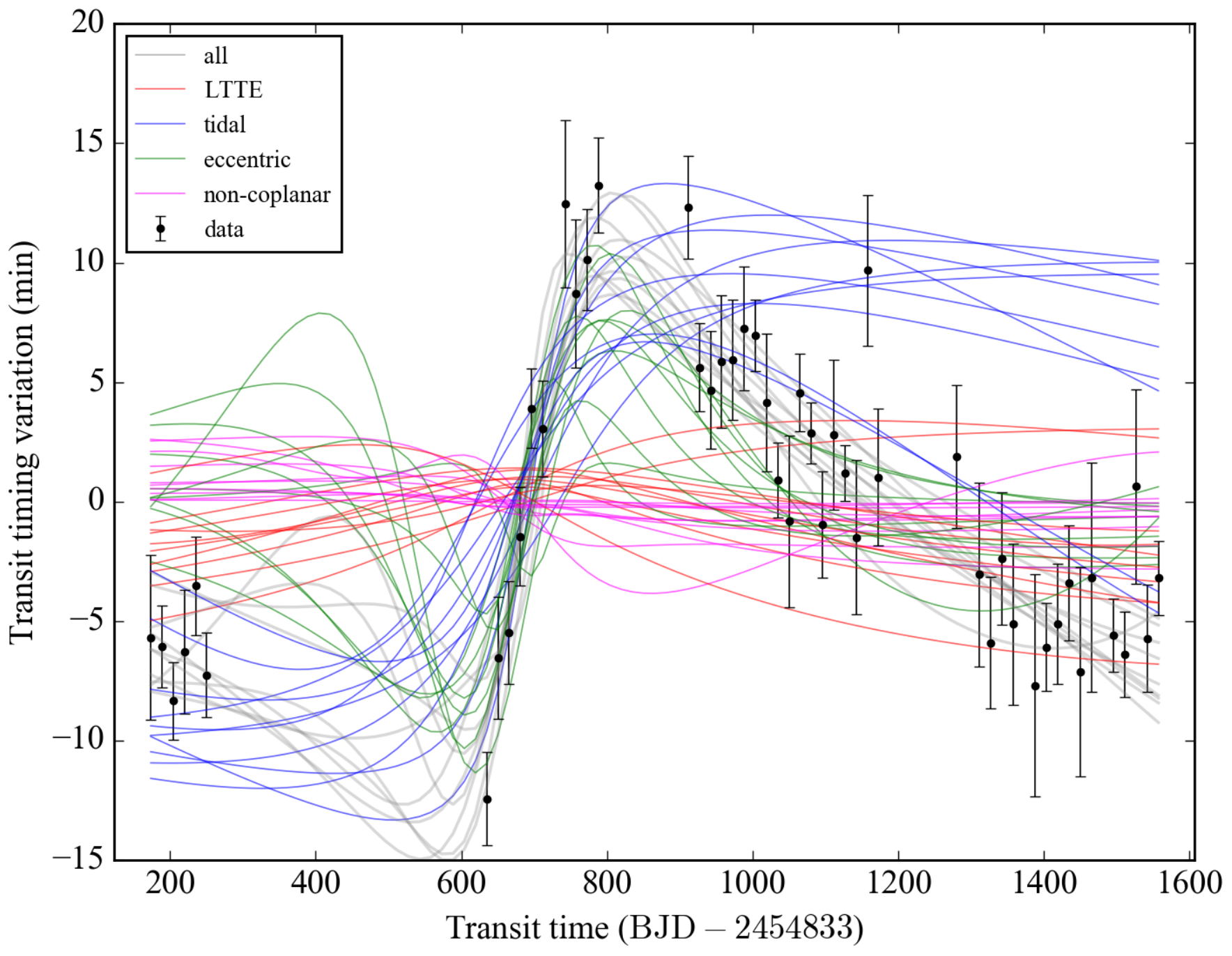}{0.49\textwidth}{(b) Kepler-693b}
	\caption{Decomposition of the analytic TTV models into individual terms in Equations \ref{eq:ltte} and \ref{eq:quad}.
	}
	\label{fig:ttv_analytic}
\end{figure*}

The analytic formula allows us to understand how each physical effect in Equation \ref{eq:quad} contributes to the observed TTVs. Figure \ref{fig:ttv_analytic} shows the decomposed signals for each of the (i) ``LTTE" $\Delta_{\rm LTTE}$, (ii) ``tidal" $\delta_{\rm tidal}$, (iii) ``eccentric" $\delta_{\rm ecc1}+\delta_{\rm ecc2}$, and (iv) ``non-coplanar" $\delta_{\rm noncopl}$ terms for $10$ solutions randomly sampled from the posterior obtained in the previous section. The plot shows that the $\delta_{\rm ecc}$ terms play a crucial role in producing the short-term feature, especially for Kepler-448b.

\section{Corner Plots for the Posteriors from Dynamical Analyses}\label{app:corner}

\begin{figure*}
	\centering
	\includegraphics[width=\textwidth]{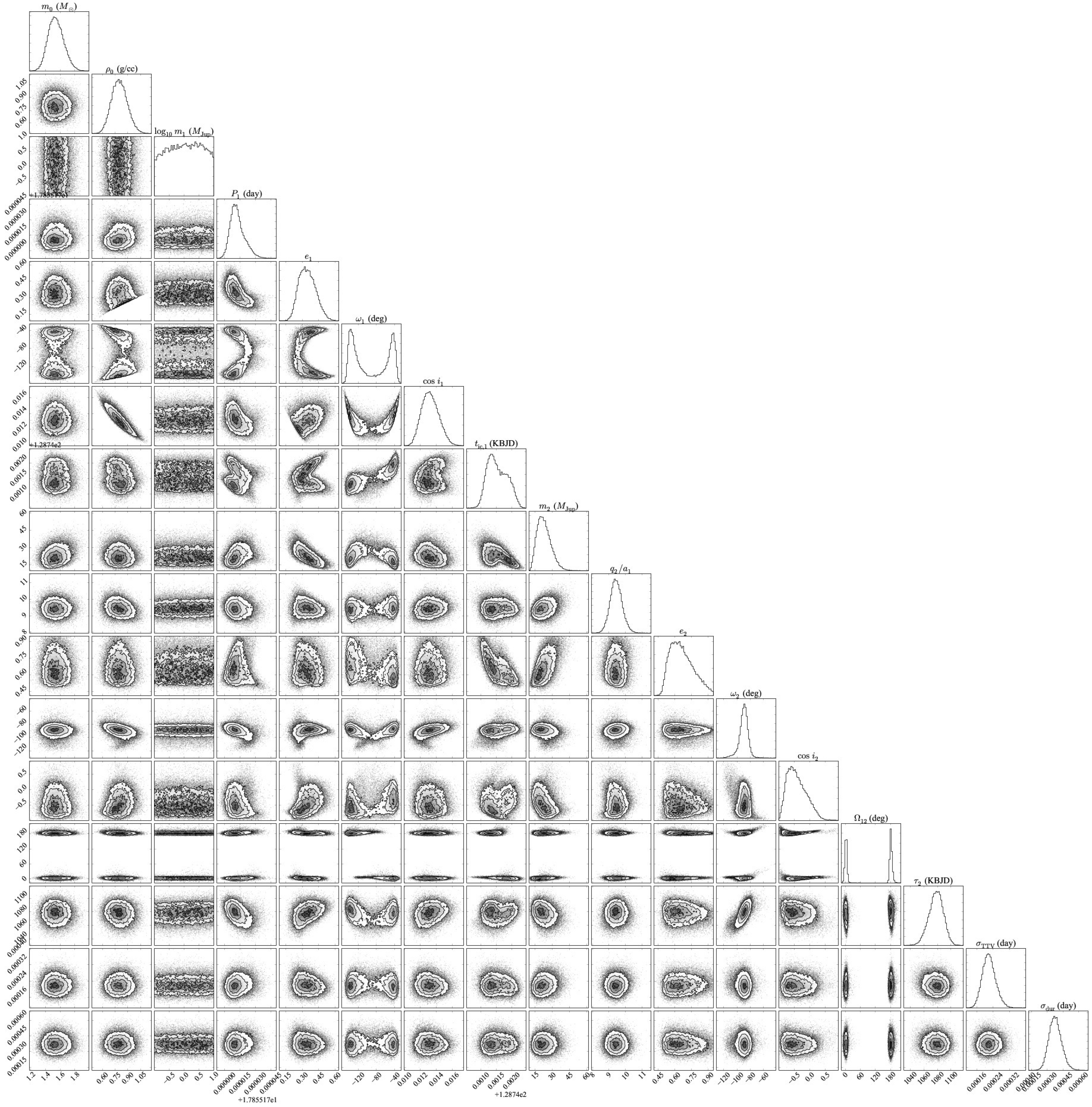}
	\caption{Joint posterior distribution from the dynamical analysis
	of Kepler-448b's TTVs and TDVs (Table \ref{tab:photod_koi12} and Figure \ref{fig:bestfit_koi12}).
	}
	\label{fig:corner_koi12}
\end{figure*}

\begin{figure*}
	\centering
	\includegraphics[width=\textwidth]{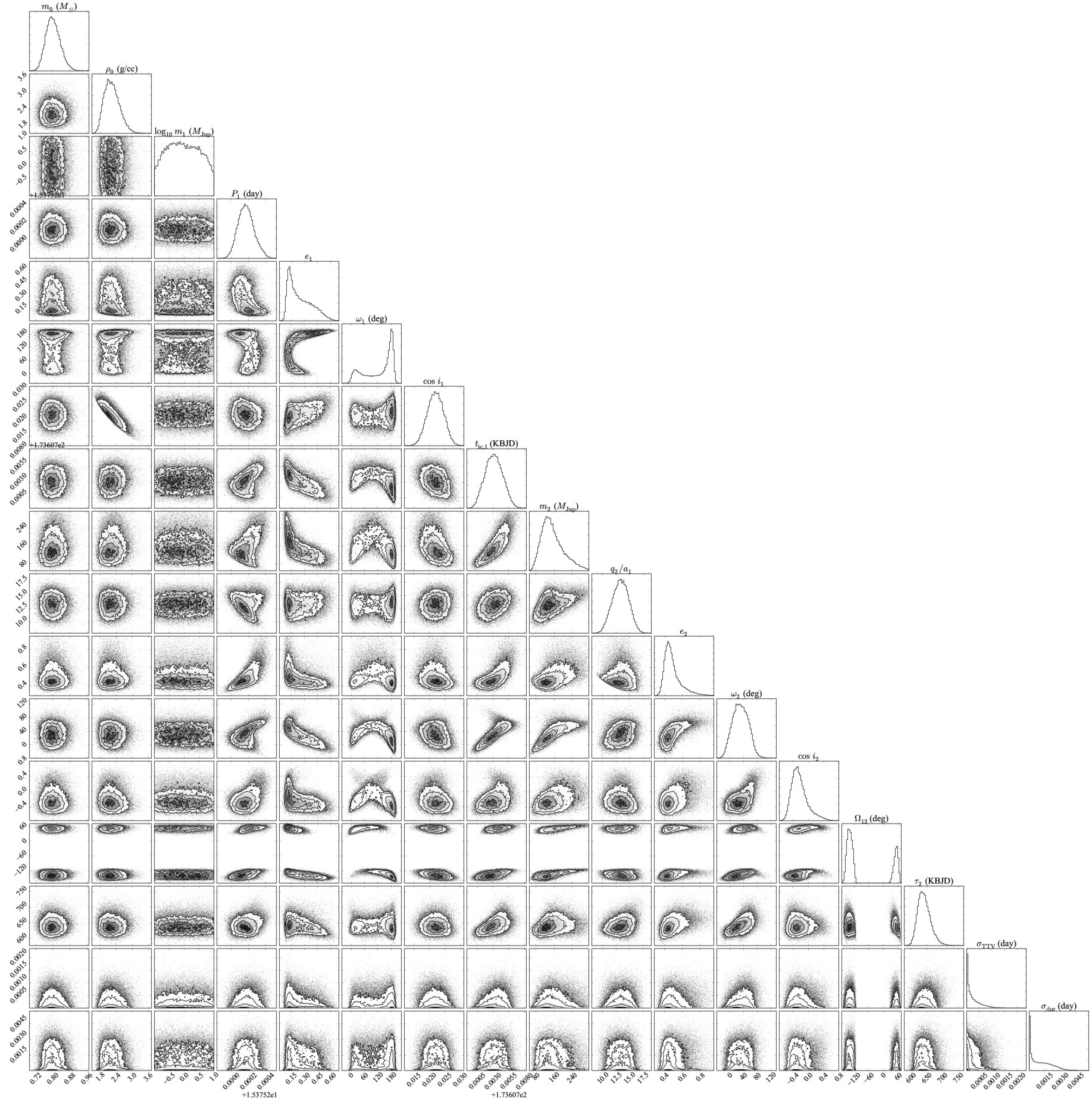}
	\caption{Joint posterior distribution from the dynamical analysis
	of Kepler-693b's TTVs and TDVs (Table \ref{tab:photod_koi824} and Figure \ref{fig:bestfit_koi824}).
	}
	\label{fig:corner_koi824}
\end{figure*}

Figures \ref{fig:corner_koi12} and \ref{fig:corner_koi824} show the corner plots of the posterior distributions obtained from the dynamical TTV and TDV analyses in Sections \ref{sec:koi12} and \ref{sec:koi824}. The figures are generated using {\tt corner.py} by \citet{corner}.

\bibliographystyle{apj}


\listofchanges

\end{document}